\documentclass[12pt, a4paper]{article}
\usepackage[latin1]{inputenc}
\usepackage{vmargin}
\usepackage{amsfonts}
\usepackage{amssymb}
\usepackage{epstopdf}
\usepackage[dvips]{graphicx}
\usepackage{graphicx}
\usepackage{amsmath}
\usepackage{amsthm}
\usepackage{setspace}
\usepackage{multirow}
\usepackage{hyperref}
\usepackage{cases}
\usepackage[bottom]{footmisc}
\usepackage[small]{caption}
\setmarginsrb{1.8cm}{1.8cm}{1.8cm}{1.8cm}{0cm}{0cm}{0cm}{0.5cm}

\begin{document}

\newtheorem{Proposition}{Proposition}
\newtheorem{Theorem}{Theorem}
\newtheorem{Corollary}{Corollary}
\newtheorem{Def}{Definition}
\newtheorem{Lem}{Lemma}

\title{Optimal Portfolio Liquidation with Limit Orders\addtocounter{footnote}{-1}\thanks{This research was partially supported by the Research Initiative ``Microstructure des marchés financiers'' under the aegis of the Europlace Institute of Finance. The authors also wish to acknowledge the helpful conversations with Yves Achdou (Université Paris-Diderot), Erhan Bayraktar (University of Michigan), Bruno Bouchard (Université Paris-Dauphine), \`Alvaro Cartea (UCL), Vincent Fardeau (LSE), Sebastian Jaimungal (University of Toronto), Jean-Michel Lasry (Université Paris-Dauphine), Mike Ludkovski (UC Santa Barbara), Gilles Pagès (Université Pierre et Marie Curie), Huyên Pham (Université Paris-Diderot) and Nizar Touzi (Ecole Polytechnique). Two anonymous referees also deserve a special thank for the improvements following their reports.}}
\author{Olivier Guéant\addtocounter{footnote}{0}\thanks{UFR de Math\'ematiques, Laboratoire Jacques-Louis Lions, Universit\'e Paris-Diderot. 175, rue du Chevaleret, 75013 Paris, France. \texttt{olivier.gueant@ann.jussieu.fr}}, Charles-Albert Lehalle\addtocounter{footnote}{5}\thanks{Head of Quantitative Research. Cr\'edit Agricole Cheuvreux. 9, Quai du Pr\'esident Paul Doumer, 92400 Courbevoie, France. \texttt{clehalle@cheuvreux.com}}, Joaquin Fernandez Tapia\addtocounter{footnote}{4}\thanks{PhD student, Université Pierre et Marie Curie. 4 place Jussieu, 75005 Paris, France.} }
\date{First draft: December 2010 -- This version: July 2012}

\maketitle
\abstract{This paper addresses portfolio liquidation using a new angle. Instead of focusing only on the scheduling aspect like Almgren and Chriss in \cite{OPTEXECAC00}, or only on the liquidity-consuming orders like Obizhaeva and Wang in \cite{ANYA05}, we link the optimal trade-schedule to the price of the limit orders that have to be sent to the limit order book to optimally liquidate a portfolio. Most practitioners address these two issues separately: they compute an optimal trading curve and they then send orders to the markets to try to follow it. The results obtained in this paper can be interpreted and used in two ways: (i) we solve simultaneously the two problems and provide a strategy to liquidate a portfolio over a few hours, (ii) we provide a tactic to follow a trading curve over slices of a few minutes. As far as the model is concerned, the interactions of limit orders with the market are modeled via a point process pegged to a diffusive ``fair price''. A Hamilton-Jacobi-Bellman equation is used to solve the control problem involving both non-execution risk and price risk. Backtests are carried out to exemplify the use of our results, both on long periods of time (for the entire liquidation process) and on slices of 5 minutes (to follow a given trading curve).
}

\section*{Introduction}

Optimal scheduling of large orders in order to control the overall trading costs with a trade-off between market impact (demanding to trade slow) and market risk (urging to trade fast) has been proposed in the litterature in the late nineties mainly by Bertsimas and Lo \cite{BLA98} and Almgren and Chriss \cite{OPTEXECAC00}.
The original approach has been recently generalized in several directions (see for instance \cite{citeulike:5177342, almgren03, NMD, citeulike:7369801, citeulike:8043820, citeulike:6699563, citeulike:5094012, lorenz2011mean, schied2009risk}); however few attempts have been made to drill down the model at the level of the interactions with the order books. The more noticeable proposal is the one by Obizhaeva and Wang \cite{ANYA05}, followed and generalized by Alfonsi, Fruth and Schied \cite{citeulike:6615020} and Predoiu, Shaikhet and  Shreve \cite{citeulike:8531791}. This branch of the optimal trading literature\footnote{see also \cite{citeulike:5797837, he2005dynamic, citeulike:2775239, kharroubi2009optimal} for different approaches.} focuses on the dynamics initiated by aggressive orders hitting a resilient order-book\footnote{See \cite{citeulike:5177397, huberman2004price} for the admissible transient market impact models.}, ignoring trading by passive orders.\\

Recall that during the continuous auction processes implemented by most electronic trading pools, market participants send their open interests (\emph{i.e.} buy or sell orders) to a queuing system where a ``first in first out'' queue stands at each possible price. If a buy (respectively sell) order reaches a queue of sell (resp. buy) orders, a transaction occurs (see for instance \cite{citeulike:7621540} for more explanations and modeling details).
Orders generating trades are said to be aggressive or liquidity-consuming ones; orders filling queues are said to be passive or liquidity-providing ones. In practice, most trading algorithms are as passive as possible (a typical balance for a scheduling algorithm is around 60\% of passively obtained trades -- see \cite{citeulike:5637814}). The economic literature first explored and studied these interactions between orders sent to a continuous auction system by different actors from a global efficiency viewpoint\footnote{see \cite{FOU06} for a study of the effect of ``smart order routing'' on competitive trading venues, or \cite{citeulike:6607108} for a study of the efficiency of order matching mechanisms.}. With the fragmentation of equity markets in the US and in Europe, the issue of linking optimal posting prices to the optimal liquidation of a portfolio is more and more important. A trading algorithm has to find an optimal scheduling or rhythm for its trading, but also to choose a sequel of prices and quantities of orders to send to the markets to follow this optimal rhythm as much as possible.\\

In contrast with most of the preceding literature\footnote{Kratz and Schöneborn \cite{citeulike:7358610} proposed an approach with both market orders and access to dark pools.}, we use a new approach, introduced in parallel by Bayraktar and Ludkovski \cite{citeulike:9272222} in a risk-neutral model, which is liquidity-providing oriented: liquidation strategies involve limit orders and not market orders. Therefore, our model may be seen as addressing two problems at the same time. It indeed solves the ``optimal scheduling and posting'' problem as a whole, but it can also be used to send limit orders within slices of a few minutes in order to follow a given trading curve.\\
Since we use limit orders, no execution cost is incurred in our model and the classical trade-off of the literature between market impact, or execution costs, and price risk disappears in our setting (except at the final time for the shares that may remain). However, since the broker does not know when his orders are going to be executed -- if at all --, a new risk is borne: (non-)execution risk. If a limit ask order is inserted in the order book, probability of execution and eventually time of execution will depend on the price of the order.\\

In our framework, the flow of trades ``hitting'' a passive order at a distance $\delta^a_t$ from a reference price (the ``fair price'') $S_t$ -- modeled by a Brownian motion -- follows an adapted point process of intensity $A\exp(-k \delta^a_t)$. It means that the further away from the ``fair price'' an order is posted, the less transactions it will obtain. In practice, if the limit order price is far above the best ask price, the trading gain may be high but execution is far from being guaranteed and the broker is exposed to the risk of a price decrease. On the contrary, if the limit order price is near the best ask price, or even reduces the market bid-ask spread, gains will be small but the probability of execution will be higher, resulting in faster trading and less price risk.\\

These modeling choices are rooted to a liquidity model that was originally introduced by Ho and Stoll \cite{HoStoll} and then modified by Avellaneda and Stoikov \cite{avst08} to model market making. The main characteristic of this model is that it does not explicitly consider the limit order book but statistically models liquidity: this is an advantage over limit order book models when it comes to mathematical tractability. Our paper, along with \cite{citeulike:9272222}, is an attempt to use this model to tackle the completely different issue of optimal liquidation. However, contrary to Bayraktar and Ludkovski \cite{citeulike:9272222} who only consider risk-neutral traders, our setting takes into account aversion to both price risk and non-execution risk.\\
Similar models have been used to deal with the initial problem of market making. Cartea, Jaimungal and Ricci \cite{cartea2011buy} consider a more sophisticated model than the Avellaneda-Stoikov one, including richer dynamics of market orders, impact on the limit order book, adverse selection effects and predictable $\alpha$, to deal with high-frequency market making. Cartea and Jaimungal \cite{cartea2012risk} recently used a similar model to introduce risk measures for high-frequency trading. Earlier, they used a model inspired from Avellaneda-Stoikov \cite{cartea2010modeling} in which the mid-price is modeled by a Hidden Markov Model. Guilbaud and Pham \cite{guilbaud2011optimal} also used a model inspired from the Avellaneda-Stoikov framework but including market orders and limit orders at best (and next to best) bid and ask together with stochastic spreads.\\
These models are all dealing with high-frequency trading and market-making but may be used for optimal liquidation issues\footnote{A recently-released working paper by Guilbaud and Pham \cite{guilbaud2012optimal} treated the two questions using a similar model in a pro-rata microstructure.}. None of them however manage to provide simple expressions for the optimal quotes\footnote{be it for market making or if they were used to deal with optimal liquidation strategies.}: they rely either on first order approximations or on numerical approximations for PDEs. This paper, by contrast, provides simple and easy-to-compute expressions for the optimal quotes when the trader is willing to liquidate a portfolio. It is also noteworthy that a companion paper \cite{citeulike:9272221} provides new analytics within the original Avellaneda-Stoikov framework and closed-form approximations for the quotes of a market maker using mathematical tools from spectral theory.\\

The remainder of this text is organized as follows. In the first section, we present the setting of the model and the main hypotheses on execution. The second section is devoted to the resolution of the partial differential equations arising from the control problem. Section 3 deals with three special cases: (i) the time-asymptotic case, (ii) the absence of price risk and the risk-neutral case, and (iii) a limiting case in which the trader has a large incentive to liquidate before the end. These special cases provide simple closed-form formulae allowing us to better understand the forces at stake. Then, in section 4, we carry out comparative statics and discuss the way optimal strategies depend on the model parameters. Finally, in section 5, we show how our approach can be used in practice for optimal liquidation, both on a long period of time, to solve the entire liquidation problem, and on slices of 5 minutes, when one wants to follow a predetermined trading curve.

\section{Setup of the model}

Let us fix a probability space $(\Omega, \mathcal{F}, \mathbb{P})$ equipped with a filtration $(\mathcal{F}_t)_{t\geq 0}$ satisfying
the usual conditions. We assume that all random variables and stochastic processes are defined on $(\Omega, \mathcal{F},(\mathcal{F}_t)_{t\geq 0}, \mathbb{P})$.\\

We consider a trader who has to liquidate a portfolio containing a large quantity $q_0$ of a given stock. We suppose that the reference price of the stock (which can be considered the mid-price or the best bid quote for example) moves as a brownian motion with a drift:
$$dS_t = \mu dt + \sigma dW_t$$
The trader under consideration will continuously propose an ask quote\footnote{In what follows, we will often call $\delta^a_t$ the quote instead of $S^a_t$.} denoted $S^a_t = S_t + \delta^a_t$ and will hence sell shares according to the rate of arrival of aggressive orders at the prices he quotes.\\
His inventory $q$, that is the quantity he holds, is given by $q_t = q_0 - N^a_t$ where $N^a$ is the jump process counting the number of shares he sold\footnote{Once the whole portfolio is liquidated, we assume that the trader remains inactive.}. We assume that jumps are of unitary size and it is important to notice that $1$ share may be understood as $1$ bunch of shares, each bunch being of the same size\footnote{Typically the average trade size (hereafter ATS) or a fraction of it. If one wants to replace orders of size $1$ by orders of size $\delta q$ in the model, this can be done easily. However, the framework of the model imposes to trade with orders of constant size, an hypothesis that is an approximation of reality since orders may in practice be partially filled.}. Arrival rates obviously depend on the price $S^a_t$ quoted by the trader and we assume that intensity $\lambda^a$ associated to $N^a$ is of the following form:\\
$$\lambda^a(\delta^{a}) = A \exp(-k\delta^a) = A \exp(-k(s^a-s))$$
This means that the lower the order price, the faster it will be executed.\\

As a consequence of his trades, the trader has an amount of cash whose dynamics is given by:\\
$$dX_t = (S_t + \delta^a_t)  dN^a_t$$

The trader has a time horizon $T$ to liquidate the portfolio and his goal is to optimize the expected utility of his P\&L at time $T$. We will focus on CARA utility functions and we suppose that the trader optimizes:\\

$$\sup_{(\delta_t^a)_t \in \mathcal{A}} \mathbb{E}\left[-\exp\left(-\gamma (X_T + q_T(S_T-b)) \right)  \right]$$
where $\mathcal{A}$ is the set of predictable processes on [0,T], bounded from below, where $\gamma$ is the absolute risk aversion characterizing the trader, where $X_T$ is the amount of cash at time $T$, where $q_T$ is the remaining quantity of shares in the inventory at time $T$ and where $b$ is a cost (per share) one has to incur to liquidate the remaining quantity at time $T$\footnote{We introduced here a constant liquidation cost but most results would be \emph{mutatis mutandis} the same for an instantaneous market impact function $b(q_T)$.}.\\

\section{Optimal quotes}

\subsection{Hamilton-Jacobi-Bellman equation}

The optimization problem set up in the preceding section can be solved using classical Bellman tools.
To this purpose, we introduce the Hamilton-Jacobi-Bellman equation associated to the optimization problem, where $u$ is an unknown function that is going to be the value function of the control problem:

$$(\mathrm{HJB}) \qquad \partial_t u(t,x,q,s) + \mu \partial_{s} u(t,x,q,s) + \frac 12 \sigma^2 \partial_{ss}^2 u(t,x,q,s)$$$$ + \sup_{\delta^a} \lambda^a(\delta^{a}) \left[u(t,x+s+\delta^{a},q-1,s) - u(t,x,q,s) \right]=0$$
with the final condition:
$$u(T,x,q,s) = -\exp\left(-\gamma (x+q(s-b)) \right) $$
and the boundary condition:
$$u(t,x,0,s) = -\exp\left(-\gamma x \right)$$

To solve the Hamilton-Jacobi-Belmann equation, we will use a change of variables that transforms the PDEs in a system of linear ODEs.

\begin{Proposition}[A system of linear ODEs]
Let us consider a family of functions $(w_q)_{q \in \mathbb{N}}$ solution of the linear system of ODEs $(\mathcal{S})$ that follows:
$$\forall q\in \mathbb{N}, \dot{w}_q(t) = (\alpha q^2 - \beta q) w_q(t) - \eta  w_{q-1}(t)$$
with $w_q(T) = e^{-kqb}$ and $w_0 = 1$, where $\alpha = \frac k2 \gamma \sigma^2$, $\beta = k \mu$ and $\eta = A (1+\frac \gamma k)^{-(1+\frac k\gamma)}$.\\

Then $u(t,x,q,s) = -\exp(-\gamma(x+qs)){w_q(t)}^{-\frac \gamma k}$ is solution of $(\mathrm{HJB})$.\\
\end{Proposition}

The change of variables used in Proposition 1 is based on two different ideas. First, the choice of a CARA utility function allows to factor out the Mark-to-Market value of the portfolio ($x+qs$). Then, the exponential decay for the intensity allows to introduce $w_q(t)$ and to end up with a linear system of ordinary differential equations.\\

Now, using this system of ODEs, we can find the optimal quotes through a verification theorem:
\begin{Theorem}[Verification theorem and optimal quotes]
Let us consider the solution $w$ of the system $(\mathcal{S})$ of Proposition 1.\\
Then, $u(t,x,q,s) = -\exp(-\gamma(x+qs)){w_q(t)}^{-\frac \gamma k}$ is the value function of the control problem and the optimal ask quote can be expressed as:
$$\delta^{a*}(t,q) = \left( \frac{1}{k}\ln\left(\frac{w_{q}(t)}{w_{q-1}(t)}\right) + \frac 1\gamma \ln\left(1+\frac \gamma k\right)\right)$$
\end{Theorem}

\subsection{Numerical example}

\begin{figure}[!h]
  \center
  \includegraphics[width=300pt]{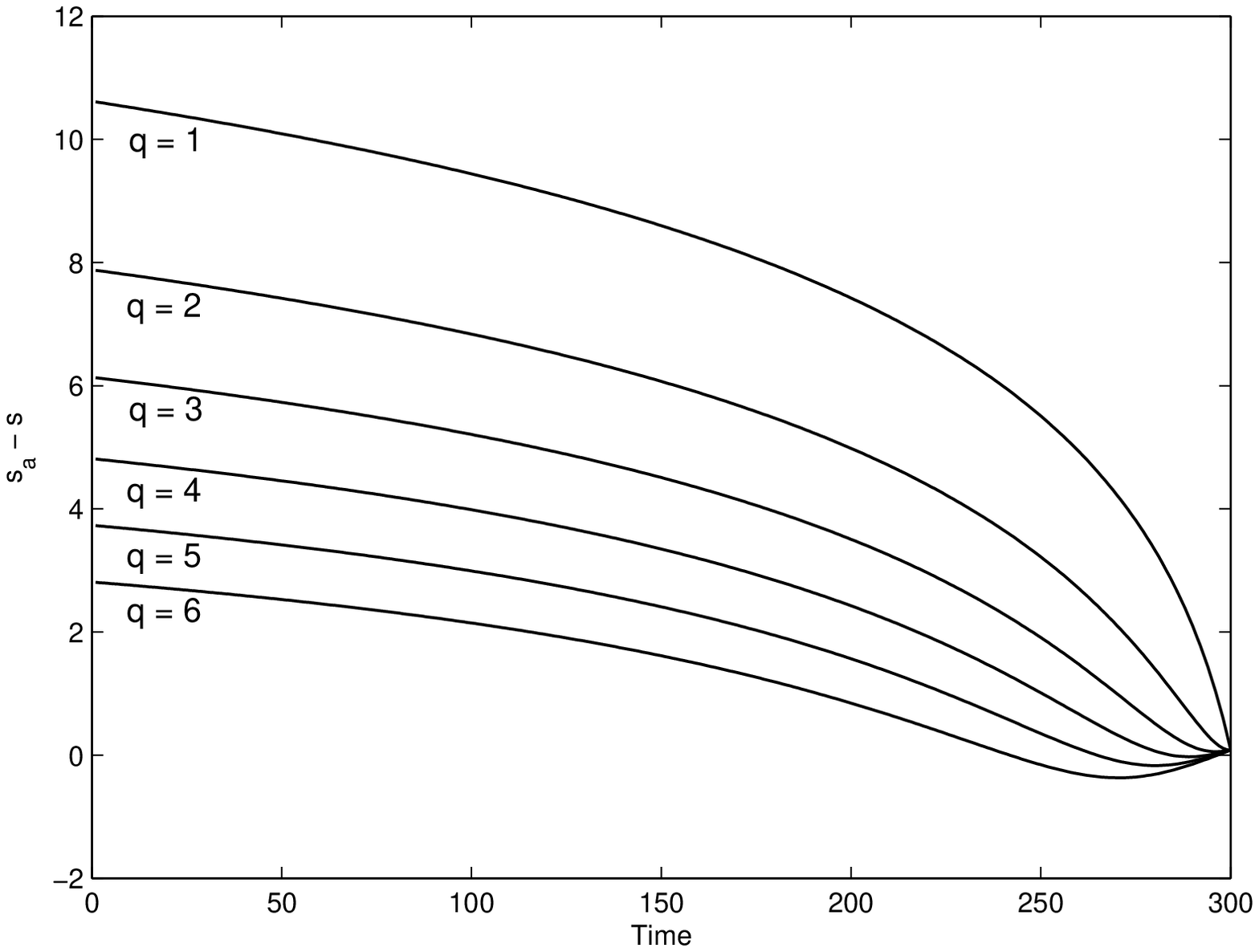}\\
  \caption{Optimal strategy $\delta^{a*}(t,q)$ (in Ticks) for an agent willing to sell a quantity up to $q=6$ within 5 minutes ($\mu = 0 \mathrm{\;} (\mathrm{ Tick}.\mathrm{s}^{-1})$, $\sigma = 0.3 \mathrm{\;} (\mathrm{ Tick}.\mathrm{s}^{-\frac{1}{2}})$, $A = 0.1 \mathrm{\;} (\mathrm{s}^{-1})$, $k = 0.3 \mathrm{\;} (\mathrm{Tick}^{-1})$, $\gamma =  0.05\mathrm{\;}(\mathrm{Tick}^{-1})$ and $b=3\mathrm{\;}(\mathrm{Tick})$)}
  \label{optstrat}
\end{figure}

Proposition 1 and Theorem 1 provide a way to solve the Hamilton-Jacobi-Bellman equation and to derive the optimal quotes for a trader willing to liquidate a portfolio. To exemplify these results, we compute the optimal quotes when a quantity $q=6$ has to be sold within 5 minutes (Figure~\ref{optstrat}).\\

We clearly see that the optimal quotes depend on inventory in a monotonic way. Indeed, a trader with a lot of shares to liquidate need to trade fast to reduce price risk and will therefore propose a low price. On the contrary a trader with only a few shares in his portfolio may be willing to benefit from a trading opportunity and will send an order with a higher price because the risk he bears allows him to trade more slowly.\\

Now, coming to the time-dependence of the quotes, a trader with a given number of shares will, \emph{ceteris paribus}, lower his quotes as the time horizon gets closer, except near the final time $T$ because a certain maximum discount $b$ is guaranteed. At the limit, when $t$ is close to the time horizon $T$, the optimal quotes tend to the same value that depend on the liquidation cost $b$: $\delta^{a*}(T,q) = -b + \frac 1\gamma \ln\left(1+\frac \gamma k\right)$.\\

As on the above figure, negative quotes may appear. They appear when the quantity to liquidate is large compared to the remaining time, especially when (i) there is a real need to liquidate before time $T$ because the liquidation cost $b$ is high and/or (ii) when risk aversion and volatility are high, because price risk is then an important consideration. When this happens, it means that there is a need to reduce the number of shares at hand. In that case, a model involving both limit orders and market orders would be better suited. In practice, one should not use a liquidation model that only involves limit orders when the volume to be executed is too large and evidently requires market orders from the very beginning. However, if a negative quote appears after some time, because of a slow execution, then sending a market order (with thus guaranteed execution) to reduce the inventory is a possible approximation of the model\footnote{It must be noted that this \emph{ad hoc} approximation ignores transaction costs associated to market orders.}.\\

\begin{figure}[!h]
  \center
  \includegraphics[width=300pt]{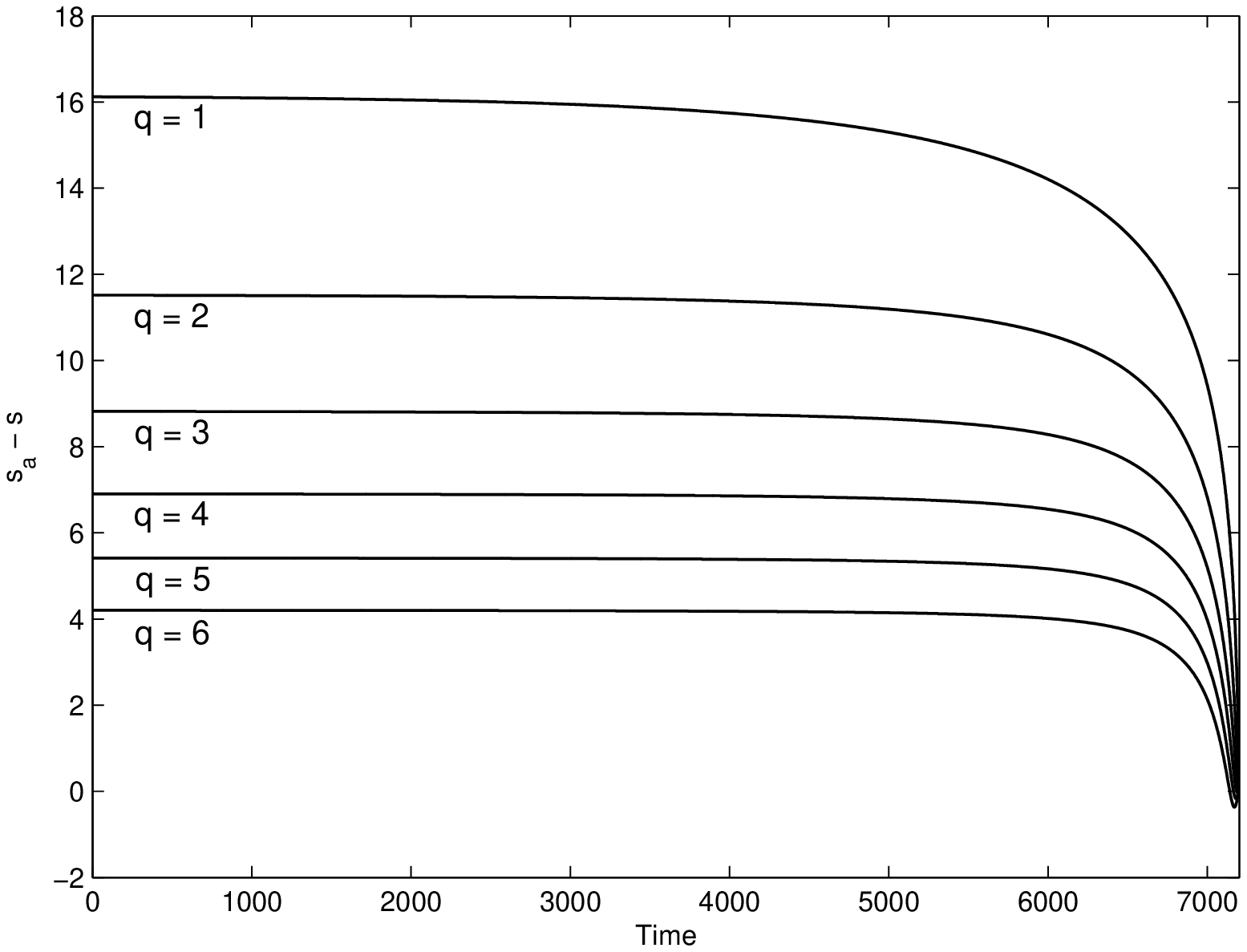}\\
  \caption{Optimal strategy $\delta^{a*}(t,q)$ (in Ticks) for $q=1, \ldots, 6$ and $T=2$ hours ($\mu = 0 \mathrm{\;} (\mathrm{ Tick}.\mathrm{s}^{-1})$, $\sigma = 0.3 \mathrm{\;} (\mathrm{ Tick}.\mathrm{s}^{-\frac{1}{2}})$, $A = 0.1 \mathrm{\;} (\mathrm{s}^{-1})$, $k = 0.3 \mathrm{\;} (\mathrm{Tick}^{-1})$, $\gamma =  0.05\mathrm{\;}(\mathrm{Tick}^{-1})$ and $b=3\mathrm{\;}(\mathrm{Tick})$)}
  \label{optstratlong}
\end{figure}

Also, if we consider the above optimal strategy on a longer time window (see Figure~\ref{optstratlong}), we see that optimal quotes have an asymptotic behavior as the time horizon increases. The associated limiting case will be studied in the next section.\\

\begin{figure}[!h]
  \center
  \includegraphics[width=300pt]{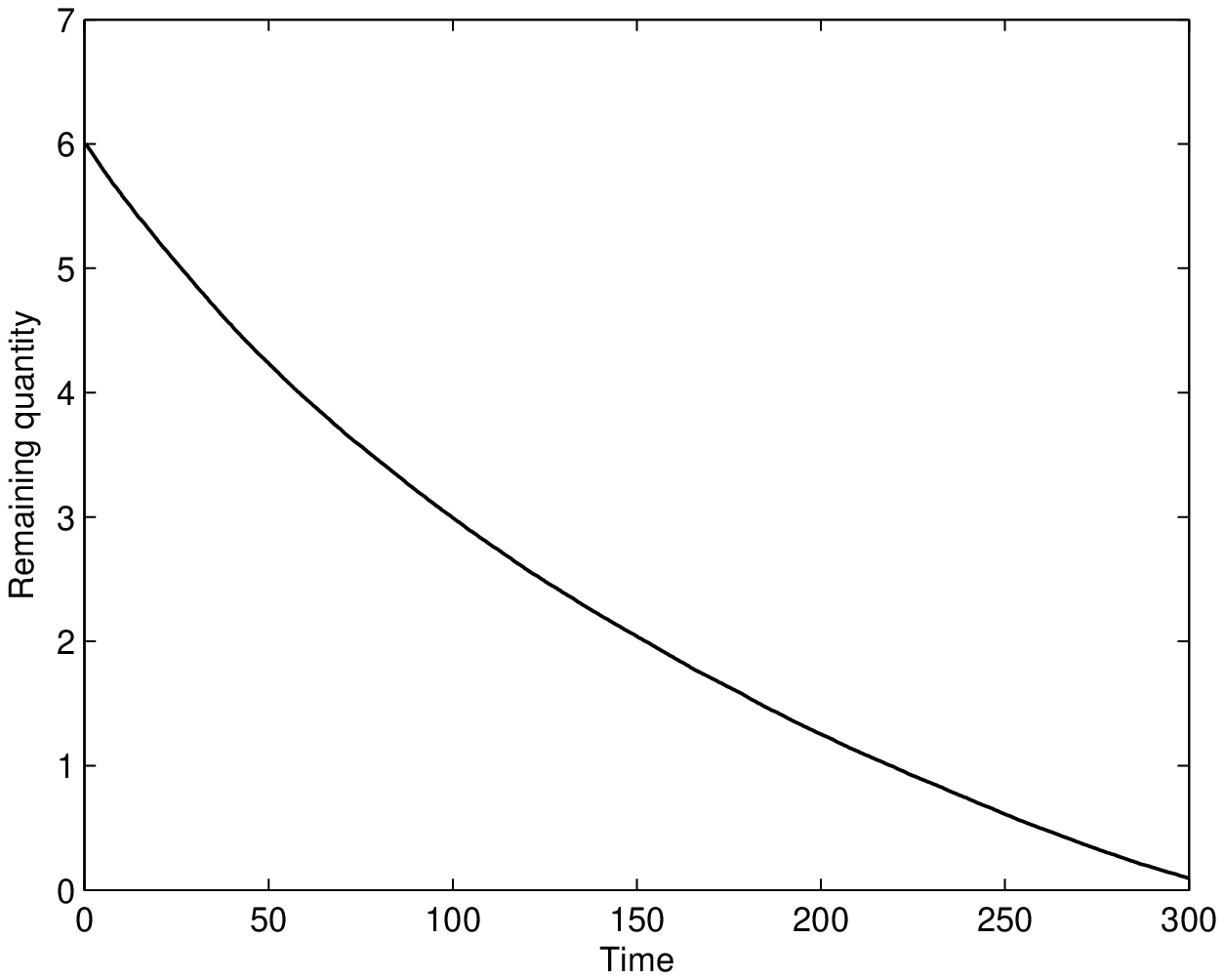}\\
  \caption{Trading curve for an agent willing to sell a quantity of shares $q=6$ within 5 minutes ($\mu = 0 \mathrm{\;} (\mathrm{ Tick}.\mathrm{s}^{-1})$, $\sigma = 0.3 \mathrm{\;} (\mathrm{ Tick}.\mathrm{s}^{-\frac{1}{2}})$, $A = 0.1 \mathrm{\;} (\mathrm{s}^{-1})$, $k = 0.3 \mathrm{\;} (\mathrm{Tick}^{-1})$, $\gamma =  0.05\mathrm{\;}(\mathrm{Tick}^{-1})$) and $b=3\mathrm{\;}(\mathrm{Tick})$}
  \label{reftradingcurve}
\end{figure}

Finally, the average number of shares at each point in time, called trading curve by analogy with the deterministic trading curves of Almgren and Chriss, can be obtained by Monte-Carlo simulations as exemplified on Figure~\ref{reftradingcurve} when the trader uses the optimal strategy.\\
\vspace{0.4cm}

In particular, because $b=3$, the trader has a weak incentive to liquidate strictly before time $T$ and there are cases for which liquidation is not complete before time $T$. This is the reason why we do not have $\mathbb{E}[q_T]=0$ on the above figure. We will study below the limiting case $b \to +\infty$ that ``forces'' liquidation before time $T$.\\

\section{Special cases}

The above equations can be solved explicitly for $w$ and hence for the optimal quotes using the above verification theorem. However, the resulting closed-form expressions are not really tractable and do not provide any intuition on the behavior of the optimal quotes. Three special cases are now considered for which simpler closed-form formulae can be derived. We start with the limiting behavior of the quotes when the time horizon $T$ tends to infinity. We then consider a case in which there is no price risk and a case where the agent is risk-neutral to both price risk and non-execution risk. We finally consider, by analogy with the classical literature, the behavior of the solution as the liquidation cost $b$ increases.\\
All these special cases allow to comment on the role of the parameters, before we carry out comparative statics in the next section.\\

\subsection{Asymptotic behavior as $T\to +\infty$}

We have seen on Figure~\ref{optstratlong} that the optimal quotes seem to exhibit an asymptotic behavior. We are in fact going to prove that $\delta^{a*}(0,q)$ tends to a limit as the time horizon $T$ tends to infinity, when the inequality $\mu < \frac 12 {\gamma \sigma^2}$ is satisfied\footnote{In particular, when $\mu=0$, this means that the result is true as soon as $\gamma > 0$ (and $\sigma>0$). As we will see below, there is no asymptotic value in the risk-neutral case.}.

\begin{Proposition}[Asymptotic behavior of the optimal quotes]
Let us suppose that\footnote{This condition is the same as $\alpha > \beta$.} $\mu < \frac 12 {\gamma \sigma^2}$.\\
Let us consider the solution $w$ of the system $(\mathcal{S})$ of Proposition 1. Then:
$$\lim_{T\to +\infty} w_q(0) = \frac{\eta^q}{q!} \prod_{j=1}^q \frac{1}{\alpha j - \beta}$$
The resulting asymptotic behavior for the optimal ask quote of Theorem 1 is:
$$\lim_{T\to +\infty} \delta^{a*}(0,q) = \frac{1}{k}\ln\left( \frac{A}{ 1+\frac{\gamma}{k}}\frac{1}{\alpha q^2 - \beta q} \right) = \frac{1}{k}\ln\left( \frac{A}{ k+\gamma}\frac{1}{\frac 12 \gamma \sigma^2 q^2 - \mu q} \right)$$
\end{Proposition}

This first closed-form formula deserves a few comments. First of all, the asymptotic quote is obviously a decreasing function of the number of shares in the portfolio. Coming to the parameters, we can analyze how the asymptotic quote depends on $\mu$, $\sigma$, $A$, $k$, and $\gamma$.\\ As $\mu$ increases, the trader increases his asymptotic quote to slow down the execution process and benefit from the price increase. As far as volatility is concerned, an increase in $\sigma$ corresponds to an increase in price risk and this provides the trader with an incentive to speed up the execution process. Therefore, it is natural that the asymptotic quote is a decreasing function of $\sigma$. Now, as $A$ increases, the asymptotic quote increases. This result is natural because if the rate of arrival of liquidity-consuming orders increases, the trader is more likely to liquidate his shares faster and posting deeper into the book allows for larger profits. Coming to $k$, the result depends on the sign of the asymptotic quote. If the asymptotic quote is positive -- this is the only interesting case since we shall not use our model when market orders are required from the start\footnote{In all cases, increasing $k$ brings the asymptotic quote closer to $0$.} --, then the asymptotic quote is a decreasing function of $k$. The mechanism at play is the same as for $A$: a decrease in $k$ increases the probability to be executed at a given price (when $\delta^{a} > 0$\footnote{The issue surrounding negative quotes is that $k \mapsto Ae^{-k\delta^a}$ is a decreasing function for $\delta^a>0$ and an increasing function for $\delta^a<0$. Subsequently, the intuition we have about $k$ in the usual case $\delta^a > 0$ is reversed for negative quotes.}) and this gives an incentive to post orders deeper into the order book. Finally, the asymptotic quote decreases as the risk aversion increases. An increase in risk aversion forces indeed the trader to reduce both price risk and non-execution risk and this leads to posting orders with lower prices.\\
One also has to notice that the asymptotic quote does not depend on the liquidation cost $b$.\\

\vspace{0.5cm}
\subsection{Absence of price risk and risk-neutrality}

The above result on asymptotic behavior does not apply when $\mu = \sigma = 0$. We now concentrate on this case in which there is no drift ($\mu=0$) and no volatility ($\sigma=0$). In this case, the trader bears no price risk because $\sigma=0$ and the only risk he faces is linked to the non-execution of his orders.\\

We now derive tractable formulae for $w$ and for the optimal quotes:

\begin{Proposition}[The no-drift/no-volatility case]
Assume that $\sigma=0$ and that there is no drift ($\mu=0$).\\
Let us define:
$$w_q(t) = \sum_{j=0}^q \frac{\eta^j}{j!} e^{-kb(q-j)}(T-t)^j$$

Then $w$ defines a solution of the system $(\mathcal{S})$ and the optimal quote is:
$$\delta^{a*}(t,q) =  - b + \frac{1}{k}\ln\left(1 + \frac{\frac{\eta^q}{q!}(T-t)^q}{\sum_{j=0}^{q-1} \frac{\eta^j}{j!} e^{-kb(q-j)}(T-t)^j}\right) + \frac 1\gamma \ln\left(1+\frac \gamma k\right)$$
\end{Proposition}

In this no-drift/no-volatility case, the optimal quote still is an increasing function of $A$ and a decreasing function of $\gamma$\footnote{Recall that $\eta = A (1+\frac \gamma k)^{-(1+\frac k\gamma)}$.}. If the above closed-form formula does not shed any particular light on the dependence on $k$, it highlights the role played by the liquidation cost $b$. Differentiating the above formula, we indeed get a negative sign and therefore that optimal quote is a decreasing function of $b$. Since $b$ is the cost to pay for each share remaining at time $T$, an increase in $b$ gives an incentive to speed up execution and hence to lower the quotes.\\

We also see that the optimal quote is bounded from below by $-b+ \frac 1\gamma \ln\left(1+\frac \gamma k\right)$. Since execution is guaranteed at price $s-b$ at time $T$, it is in particular natural in the absence of price risk, that quotes never go below $-b$.\\

Now, if one wants to remove risk aversion with respect to both price risk and non-execution risk, one can consider the limit of the above solution when $\gamma$ tends to $0$\footnote{The same result holds if one sends $\gamma$ to $0$ for any value of the volatility parameter $\sigma$.}.\\

One then obtains: $$\delta^{a*}(t,q) = -b + \frac{1}{k}\ln\left(1 + \frac{\frac{A^q}{e^{q}q!}(T-t)^q}{\sum_{j=0}^{q-1} \frac{A^j }{e^{j}j!} e^{-kb(q-j)}(T-t)^j}\right) + \frac 1k $$
and this is the result of Bayraktar and Ludkovski \cite{citeulike:9272222} in the case $b=0$, because they do not consider any liquidation cost. In particular, the optimal quote of \cite{citeulike:9272222} does not converge to a limit value as $T$ tends to $+\infty$,  but rather increases with no upper bound. This is an important difference between the risk-neutral case and our risk-adverse framework.\\

\subsection{Limiting behavior as $b\to +\infty $}

Let us now consider the limiting case $b \to +\infty$. Sending $b$ to infinity corresponds to a situation in which a very high incentive is given to the trader for complete liquidation before time $T$. If we look at the Almgren-Chriss-like literature on optimal execution, the authors are often assuming that $q_T=0$\footnote{The authors most often consider target problems in which the target can always be attained.}. Hence, if one writes the value functions associated to most liquidity-consuming optimal strategies, it turns out that they are equal to $-\infty$ at the time horizon $T$ except when the inventory is equal to nought (hence $b=+\infty$, in our framework). However, here, due to the uncertainty on execution, we cannot write a well-defined control problem when $b$ is equal to $+\infty$. Rather, we are interested in the limiting behavior when $b\to +\infty$, \emph{i.e.} when the incentive to liquidate before time $T$ is large.\\

By analogy with the initial literature on optimal liquidation \cite{OPTEXECAC00}, we can also have some limiting results on the trading curve.\\

Hereafter we denote $w_{b,q}(t)$ the solution of the system $(\mathcal{S})$ for a given liquidation cost $b$, $\delta_b^{a*}(t,q)$ the associated optimal quote and $q_{b,t}$ the resulting process modeling the number of stocks in the portfolio.\\

\begin{Proposition}[Form of the solutions, trading intensity and trading curve]
The limiting solution $\lim_{b\to +\infty} w_{b,q}(t)$ is of the form $A^q v_q(t)$ where $v$ does not depend on $A$.\\

The limit of the trading intensity $\lim_{b\to +\infty} A e^{-k\delta_b^{a*}}$ does not depend on $A$.\\

Consequently, the trading curve $V_b(t) := \mathbb{E}[q_{b,t}]$ verifies that $V(t) = \lim_{b\to +\infty} V_b(t)$ is independent of $A$.
\end{Proposition}

More results can be obtained in the no-volatility case:

\begin{Proposition}[no-volatility case, $b \to +\infty$]
Assume that $\sigma=0$ and consider first the case $\mu \neq 0$. We have:\\

$$\lim_{b\to+\infty}w_{b,q}(t) = \frac{\eta^q}{q!} \left(\frac{e^{\beta(T-t)} - 1}{\beta}\right)^q$$
The limit of the optimal quote is:
$$\delta_\infty^{a*}(t,q)= \lim_{b \to +\infty} \delta_b^{a*}(t,q) = \frac{1}{k}\ln\left( \frac{A}{ 1+\frac{\gamma}{k}  }\frac{1}{q}\frac{e^{\beta(T-t)} - 1}{\beta} \right)$$
The limit of the associated trading curve is $V(t) = q_0 \left(\frac{1-e^{-\beta (T-t)}}{1-e^{-\beta T}}\right)^{1+\frac{\gamma}{k}}$.\\

Now, in the no-volatility/no-drift case ($\sigma=\mu=0$), similar results can be obtained, either directly or sending $\mu$ to $0$ in the above formulae:\\
$$\lim_{b\to+\infty}w_{b,q}(t) = \frac{\eta^q}{q!} (T-t)^q$$
The limit of the optimal quote is given by:
$$ \delta_\infty^{a*}(t,q) =\lim_{b \to +\infty} \delta_b^{a*}(t,q) = \frac{1}{k}\ln\left( \frac{A}{ 1+\frac{\gamma}{k}  }\frac{1}{q}(T-t) \right)$$
The limit of the associated trading curve is $V(t) = q_0 \left(1-\frac tT\right)^{1+\frac{\gamma}{k}}$.\\
\end{Proposition}

This third limiting case confirms the monotonicity results we discussed above: the optimal quote is an increasing function of $A$ and $\mu$\footnote{Recall that $\beta=k\mu$.} and it is a decreasing function of $\gamma$ (and of the number of shares). Concerning the shape of the trading curve, the role played by the risk aversion parameter $\gamma$ is the same as in Almgren-Chriss: an increase in $\gamma$ forces the trader to speed up the execution process and therefore steepens the slope of the trading curve. The role of $\mu$ is also interesting because a positive trend goes against the naturally convex shape of the trading curve. Since a trader slows down the execution process to benefit from a positive trend, there is a trade-off between positive trend on one hand and price risk on the other, and the trading curve may turn out to be concave when the upward trend is sufficiently important to compensate the effect of risk aversion.\\

Coming to $k$, this third limiting case is particularly interesting because there is no lower bound to the optimal quotes and we have seen above that the occurrence of negative quotes was a problem to interpret the parameter $k$. Hence, the limiting case $b \to +\infty$ appears to be a worst case.\\

In normal circumstances, we expect the optimal quote to be a decreasing function of $k$. However, straightforward computations give (in the no-drift case) that $\frac{d\delta_\infty^{a*}(t,q)}{dk} = -\frac{1}{k} \delta_\infty^{a*}(t,q) + \frac{1}{k^2} \frac{\gamma}{\gamma + k}$. The sign of this expression being negative if and only if $\delta_\infty^{a*}(t,q)$ is above a certain positive threshold, the dependence on $k$ may be reversed even for positive (but low) quotes. In the case of the asymptotic (and constant) quote discussed above, the threshold was $0$. Here, in the dynamic case under consideration, the high probability of negative optimal quotes in the future may break the monotonicity on $k$ and that is the reason why the threshold is positive.\\
Although this limiting case is rather extreme, it illustrates well the issues of the model when execution is too slow and would ideally require market orders. It is noteworthy that in the comparative statics we carry out in the next section, the usual monotonicity property is only broken for extreme values of the parameters. Also, in most reasonable cases we considered in practice, the quotes were decreasing in $k$.\\

\section{Comparative statics}

We discussed above the role played by the different parameters in particular limiting cases. We now consider the general case and carry out comparative statics on optimal quotes. The tables we obtain confirm the intuitions we developed in the preceding section.\\
$$$$
\emph{\underline{Influence of the drift $\mu$:}}\\

As far as the drift is concerned, quotes are naturally increasing with $\mu$. If indeed the trader expects the price to move down, he is going to send orders at low prices to be executed fast and to reduce the impact of the decrease in price on the P\&L. On the contrary, if he expects the price to rise, he is going to post orders deeper in the book in order to slow down execution and benefit from the price increase. This is well exemplified by Table~\ref{dep_mu}.\\

\begin{table}[!h]
  \center
  \begin{tabular}{|c||c|c|c|}
\hline
q & $\mu = -0.01 \mathrm{\;} (\mathrm{ Tick}.\mathrm{s}^{-1})$ & $\mu = 0 \mathrm{\;} (\mathrm{ Tick}.\mathrm{s}^{-1})$ & $\mu = 0.01 \mathrm{\;} (\mathrm{ Tick}.\mathrm{s}^{-1})$ \\
\hline
\hline
1 & 9.2252 & 10.6095 & 12.2329 \\
\hline
2 & 6.581 & 7.8737 & 9.3921 \\
\hline
3 & 4.92 & 6.1299 & 7.5507 \\
\hline
4 & 3.6732 & 4.8082 & 6.1391 \\
\hline
5 & 2.6607 & 3.728 & 4.9765 \\
\hline
6 & 1.8012 & 2.8073 & 3.9806 \\
\hline
\end{tabular}
  \caption{Dependence on $\mu$ of $\delta^{a*}(0,q)$ with $T= 5 \mathrm{\;} (\mathrm{minutes})$, $\sigma = 0.3\mathrm{\;} (\mathrm{Tick}.\mathrm{s}^{-\frac{1}{2}})$, $A = 0.1\mathrm{\;} (\mathrm{s}^{-1})$, $k = 0.3\mathrm{\;} (\mathrm{Tick}^{-1})$, $\gamma = 0.05\mathrm{\;} (\mathrm{Tick}^{-1})$ and $b=3 \mathrm{\;} (\mathrm{Tick})$}
  \label{dep_mu}
\end{table}\vspace{0.8cm}

\emph{\underline{Influence of the volatility $\sigma$:}}\\

Now, coming to volatility, the optimal quotes depend on $\sigma$ in a monotonic way. If there is an increase in volatility, then price risk increases. In order to reduce this additional price risk the trader will send orders at lower price. This is what we observe numerically on Table~\ref{dep_sigma}.\\

\begin{table}[!h]
  \center
\begin{tabular}{|c||c|c|c|}
\hline
q & $\sigma = 0\mathrm{\;} (\mathrm{Tick}.\mathrm{s}^{-\frac{1}{2}})$ & $\sigma = 0.3\mathrm{\;} (\mathrm{Tick}.\mathrm{s}^{-\frac{1}{2}})$ & $\sigma = 0.6\mathrm{\;} (\mathrm{Tick}.\mathrm{s}^{-\frac{1}{2}})$ \\
\hline
\hline
1 & 10.9538 & 10.6095 & 9.6493 \\
\hline
2 & 8.6482 & 7.8737 & 6.0262 \\
\hline
3 & 7.3019 & 6.1299 & 3.6874 \\
\hline
4 & 6.3486 & 4.8082 & 1.9455 \\
\hline
5 & 5.6109 & 3.728 & 0.55671 \\
\hline
6 & 5.0097 & 2.8073 & -0.59773 \\
\hline
\end{tabular}
  \caption{Dependence on $\sigma$ of $\delta^{a*}(0,q)$ with $T= 5 \mathrm{\;} (\mathrm{minutes})$, $\mu = 0\mathrm{\;} (\mathrm{Tick}.\mathrm{s}^{-1})$, $A = 0.1\mathrm{\;} (\mathrm{s}^{-1})$, $k = 0.3\mathrm{\;} (\mathrm{Tick}^{-1})$, $\gamma = 0.05\mathrm{\;} (\mathrm{Tick}^{-1})$ and $b=3 \mathrm{\;} (\mathrm{Tick})$}
  \label{dep_sigma}
\end{table}\vspace{0.8cm}

\emph{\underline{Influence of the intensity scale parameter $A$:}}\\

Now, coming to $A$, we observe numerically, and as expected, that the optimal quote is an increasing function of $A$ (see Table~\ref{dep_A}). If $A$ increases, the probability to be executed indeed increases and the trader will then increase his quotes to obtain transactions at higher prices.\\

\begin{table}[!h]
  \center
\begin{tabular}{|c||c|c|c|}
\hline
q & $A = 0.05\mathrm{\;} (\mathrm{s}^{-1})$ & $A = 0.1\mathrm{\;} (\mathrm{s}^{-1})$ & $A = 0.15\mathrm{\;} (\mathrm{s}^{-1})$ \\
\hline
\hline
1 & 8.4128 & 10.6095 & 11.9222 \\
\hline
2 & 5.6704 & 7.8737 & 9.1898 \\
\hline
3 & 3.9199 & 6.1299 & 7.4491 \\
\hline
4 & 2.5917 & 4.8082 & 6.1302 \\
\hline
5 & 1.5051 & 3.728 & 5.0525 \\
\hline
6 & 0.57851 & 2.8073 & 4.1341 \\
\hline
\end{tabular}
  \caption{Dependence on $A$ of $\delta^{a*}(0,q)$ with $T= 5 \mathrm{\;} (\mathrm{minutes})$, $\mu = 0\mathrm{\;} (\mathrm{Tick}.\mathrm{s}^{-1})$, $\sigma = 0.3\mathrm{\;} (\mathrm{Tick}.\mathrm{s}^{-\frac{1}{2}})$, $k = 0.3\mathrm{\;} (\mathrm{Tick}^{-1})$, $\gamma = 0.05\mathrm{\;} (\mathrm{Tick}^{-1})$ and $b=3 \mathrm{\;} (\mathrm{Tick})$}
  \label{dep_A}
\end{table}\vspace{0.8cm}

\emph{\underline{Influence of the intensity shape parameter $k$:}}\\

Now, as far as $k$ is concerned, the dependence of the optimal quote on $k$ is ambiguous because the interpretation of $k$ depends on the optimal quote itself. An increase in $k$ should correspond indeed to a decrease in the probability to be executed at a given price in most cases the model is used. However, due to the exponential form of the execution intensity, the very possibility to use negative quotes may reverse the reasoning (see the discussions in section 3 for the asymptotic quotes and in the extreme case $b \to +\infty$).\\

In the first case we consider, which only leads to positive optimal quotes, an increase in $k$ forces the trader to decrease the price of the orders he sends to the market, as exemplified by Table~\ref{dep_k}. However, if price risk is really important (high volatility and/or large quantity to liquidate) the optimal quotes may be negative and, in that case, the role of $k$ is reversed. This is the case when $\sigma$ takes (unrealistically) high values, as exemplified on Table~\ref{dep_k2}.\\

\begin{table}[!h]
  \center
  \begin{tabular}{|c||c|c|c|}
\hline
q & $k = 0.2 \mathrm{\;} (\mathrm{Tick}^{-1})$ & $k = 0.3 \mathrm{\;} (\mathrm{Tick}^{-1})$ & $k = 0.4 \mathrm{\;} (\mathrm{Tick}^{-1})$ \\
\hline
\hline
1 & 15.8107 & 10.6095 & 7.941 \\
\hline
2 & 11.9076 & 7.8737 & 5.7972 \\
\hline
3 & 9.4656 & 6.1299 & 4.4144 \\
\hline
4 & 7.6334 & 4.8082 & 3.3618 \\
\hline
5 & 6.1436 & 3.728 & 2.5011 \\
\hline
6 & 4.8761 & 2.8073 & 1.7688 \\
\hline
\end{tabular}
  \caption{Dependence on $k$ of $\delta^{a*}(0,q)$ with $T= 5 \mathrm{\;} (\mathrm{minutes})$, $\mu = 0\mathrm{\;} (\mathrm{Tick}.\mathrm{s}^{-1})$, $\sigma = 0.3\mathrm{\;} (\mathrm{Tick}.\mathrm{s}^{-\frac{1}{2}})$, $A = 0.1\mathrm{\;} (\mathrm{s}^{-1})$, $\gamma = 0.05\mathrm{\;} (\mathrm{Tick}^{-1})$ and $b=3 \mathrm{\;} (\mathrm{Tick})$}
  \label{dep_k}
\end{table}\vspace{0.8cm}
\begin{table}[!h]
  \center
  \begin{tabular}{|c||c|c|c|}
\hline
q & $k = 0.2 \mathrm{\;} (\mathrm{Tick}^{-1})$ & $k = 0.3 \mathrm{\;} (\mathrm{Tick}^{-1})$ & $k = 0.4 \mathrm{\;} (\mathrm{Tick}^{-1})$ \\
\hline
\hline
1 & 2.8768 & 0.79631 & -0.031056 \\
\hline
2 & -4.0547 & -3.8247 & -3.4968 \\
\hline
3 & -8.1093 & -6.5278 & -5.5241 \\
\hline
4 & -10.9861 & -8.4457 & -6.9625 \\
\hline
5 & -13.2176 & -9.9333 & -8.0782 \\
\hline
6 & -15.0408 & -11.1488 & -8.9899 \\
\hline
\end{tabular}
  \caption{Dependence on $k$ of $\delta^{a*}(0,q)$ with $T= 5 \mathrm{\;} (\mathrm{minutes})$, $\mu = 0\mathrm{\;} (\mathrm{Tick}.\mathrm{s}^{-1})$, $\sigma = 3\mathrm{\;} (\mathrm{Tick}.\mathrm{s}^{-\frac{1}{2}})$, $A = 0.1\mathrm{\;} (\mathrm{s}^{-1})$, $\gamma = 0.05\mathrm{\;} (\mathrm{Tick}^{-1})$ and $b=3 \mathrm{\;} (\mathrm{Tick})$}
  \label{dep_k2}
\end{table}\vspace{0.8cm}

\newpage

\emph{\underline{Influence of the risk aversion $\gamma$:}}\\

Turning to the risk aversion parameter $\gamma$, two effects are at stake that go in the same direction. The risk aversion is indeed common for both price risk and non-execution risk. Hence if risk aversion increases, the trader will try to reduce both price risk and non-execution risk, thus selling at lower price. We indeed see on Table~\ref{dep_gamma} that optimal quotes are decreasing in $\gamma$.\\

\begin{table}[!h]
  \center
  \begin{tabular}{|c||c|c|c|}
\hline
q & $\gamma = 0.01 \mathrm{\;} (\mathrm{Tick}^{-1})$ & $\gamma = 0.05 \mathrm{\;} (\mathrm{Tick}^{-1})$ & $\gamma = 0.5 \mathrm{\;} (\mathrm{Tick}^{-1})$ \\
\hline
\hline
1 & 11.2809 & 10.6095 & 9.84 \\
\hline
2 & 8.8826 & 7.8737 & 6.7461 \\
\hline
3 & 7.4447 & 6.1299 & 4.7262 \\
\hline
4 & 6.4008 & 4.8082 & 3.189 \\
\hline
5 & 5.5735 & 3.728 & 1.9384 \\
\hline
6 & 4.8835 & 2.8073 & 0.88139 \\
\hline
\end{tabular}
  \caption{Dependence on $\gamma$ of $\delta^{a*}(0,q)$ with $T= 5 \mathrm{\;} (\mathrm{minutes})$, $\mu = 0\mathrm{\;} (\mathrm{Tick}.\mathrm{s}^{-1})$, $\sigma = 3\mathrm{\;} (\mathrm{Tick}.\mathrm{s}^{-\frac{1}{2}})$, $A = 0.1\mathrm{\;} (\mathrm{s}^{-1})$, $k = 0.3\mathrm{\;} (\mathrm{Tick}^{-1})$, and $b=3 \mathrm{\;} (\mathrm{Tick})$}
  \label{dep_gamma}
\end{table}\vspace{0.3cm}

The above table highlights the importance of risk aversion for the optimal liquidation problem. When the number of shares to liquidate is not too small, we indeed see that the optimal quotes depend strongly on $\gamma$. In particular, our reference case $\gamma=0.05$ is really different from the case $\gamma=0.01$ and therefore very different from the risk-neutral case of \cite{citeulike:9272222}.\\

\emph{\underline{Influence of the liquidation cost $b$:}}\\

Finally, the influence of the liquidation cost $b$ is straightforward. If $b$ increases, then the need to sell strictly before time $T$ is increased because the value of any remaining share at time $T$ decreases. Hence, the optimal quotes must be decreasing in $b$ and this is what we observe on Table~\ref{dep_b}.\\

\begin{table}[!h]
  \center
\begin{tabular}{|c||c|c|c|}
\hline
q & $b = 0\mathrm{\;} (\mathrm{Tick})$ & $b = 3\mathrm{\;} (\mathrm{Tick})$ & $b = 20\mathrm{\;} (\mathrm{Tick})$ \\
\hline
\hline
1 & 10.7743 & 10.6095 & 10.4924 \\
\hline
2 & 8.0304 & 7.8737 & 7.7685 \\
\hline
3 & 6.278 & 6.1299 & 6.0353 \\
\hline
4 & 4.9477 & 4.8082 & 4.7229 \\
\hline
5 & 3.859 & 3.728 & 3.6509 \\
\hline
6 & 2.9301 & 2.8073 & 2.7374 \\
\hline
\end{tabular}
  \caption{Dependence on $b$ of $\delta^{a*}(0,q)$ with $T= 5 \mathrm{\;} (\mathrm{minutes})$, $\mu = 0\mathrm{\;} (\mathrm{Tick}.\mathrm{s}^{-1})$, $\sigma = 3\mathrm{\;} (\mathrm{Tick}.\mathrm{s}^{-\frac{1}{2}})$, $A = 0.1\mathrm{\;} (\mathrm{s}^{-1})$, $k = 0.3\mathrm{\;} (\mathrm{Tick}^{-1})$ and $\gamma = 0.05\mathrm{\;} (\mathrm{Tick}^{-1})$}
  \label{dep_b}
\end{table}

\section{Historical simulations}

Before using the above model in reality, we need to discuss some features of the model that need to be adapted before any backtest is possible.\\

First of all, the model is continuous in both time and space while the real control problem under consideration is intrinsically discrete in space, because of the tick size, and discrete in time, because orders have a certain priority and changing position too often reduces the actual chance to be reached by a market order. Hence, the model has to be reinterpreted in a discrete way. In terms of prices, quotes must not be between two ticks and we decided to round the optimal quotes to the nearest tick\footnote{We also, alternatively, randomized the choice with probabilities that depend on the respective proximity to the neighboring quotes.}. In terms of time, an order is sent to the market and is not canceled nor modified for a given period of time $\Delta t$, unless a trade occurs and, though perhaps partially, fills the order. Now, when a trade occurs and changes the inventory or when an order stayed in the order book for longer than $\Delta t$, then the optimal quote is updated and, if necessary, a new order is inserted.\\

Now, concerning the parameters, $\sigma$, $A$ and $k$ can be calibrated on trade-by-trade limit order book data while $\gamma$ has to be chosen. However, it is well known by practitioners that $A$ and $k$ have to depend at least on the market bid-ask spread. Since we do not explicitly take into account the underlying market, there is no market bid-ask spread in the model. Thus, we simply chose to calibrate\footnote{$A$ and $k$ are updated as new information comes. Interestingly, Cartea, Jaimungal and Ricci \cite{cartea2011buy} considered stochastic parameters and manage to derive a first order approximation for the optimal quotes in the case of a market making model.} $A$ and $k$ as functions of the market bid-ask spread, making then an off-model hypothesis.\\
As far as $\gamma$ is concerned, a choice based on a Value at Risk limit is possible but requires the use of Monte-Carlo simulations. We decided in our backtests to assign $\gamma$ a value that makes the first quote $\delta^{a*}$ equal to $1$ for typical values of $A$ and $k$.\\

Turning to the backtests, they were carried out with trade-by-trade data and we assumed that our orders were entirely filled when a trade occurred at or above the ask price quoted by the agent. Our goal here is just to provide examples in various situations and, to exemplify the practical use of this model, we carried out several backtests\footnote{No drift in prices is assumed in the strategy used for backtesting.} on the French stock AXA, either on very short periods (slices of 5 minutes) or on slightly longer periods of a few hours. Armed with our experience of the model, we believe that it is particularly suited to optimize liquidation within slices of a global trading curve, be it a TWAP, a VWAP, or an Implementation Shortfall trading curve.\\

\begin{figure}[!h]
  \center
  \includegraphics[width=320pt]{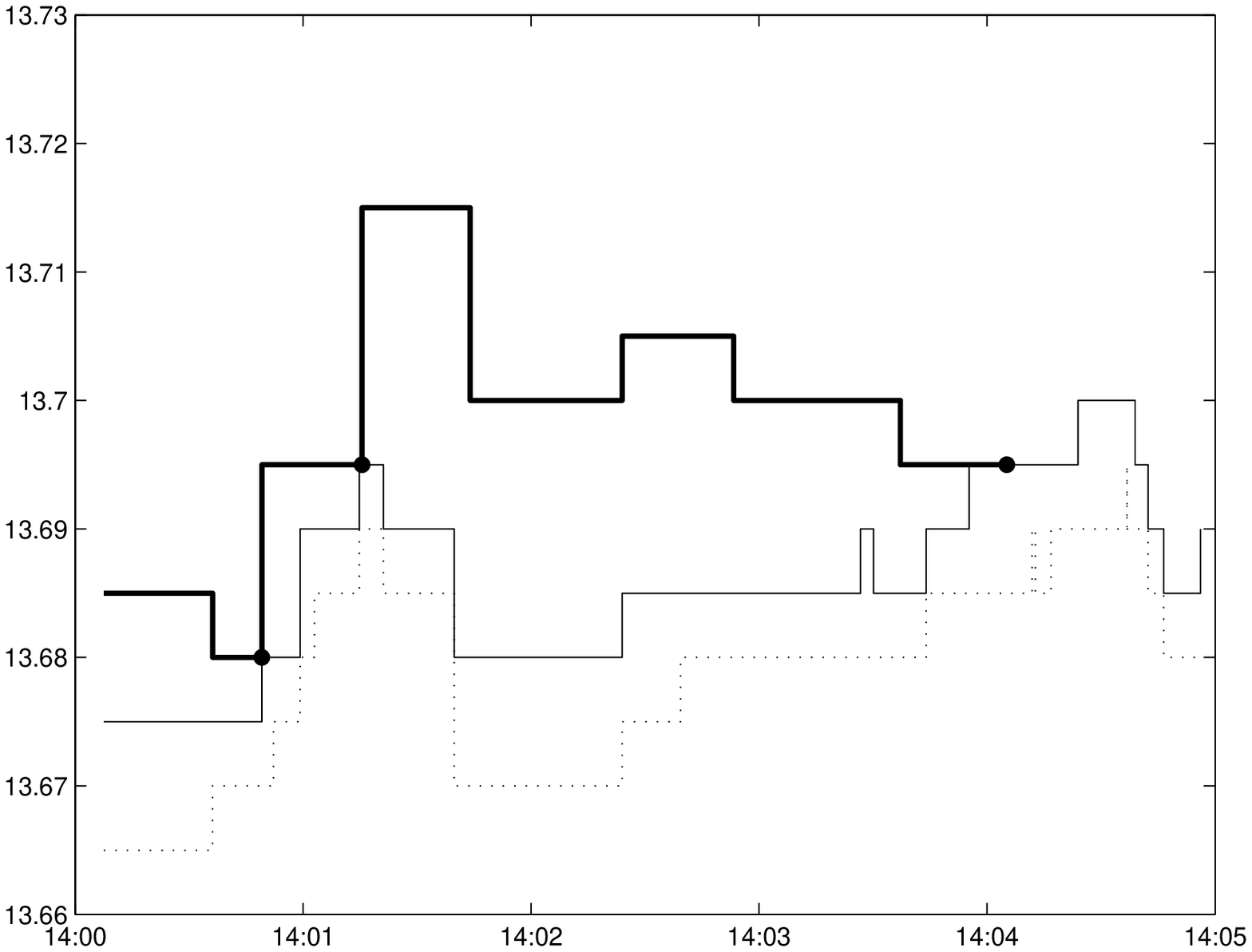}\\

  \includegraphics[width=180pt]{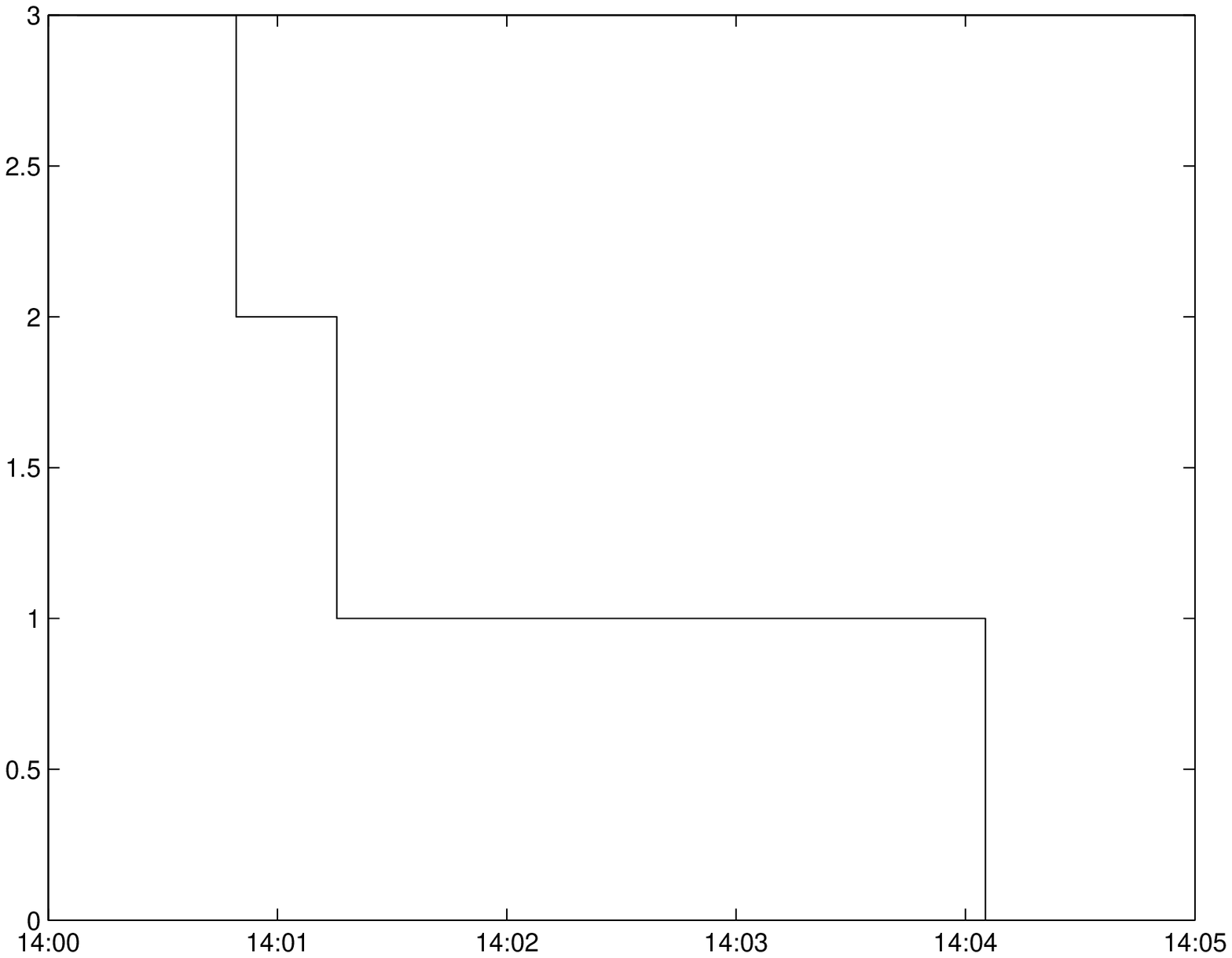}
  \hspace{1cm}
  \includegraphics[width=180pt]{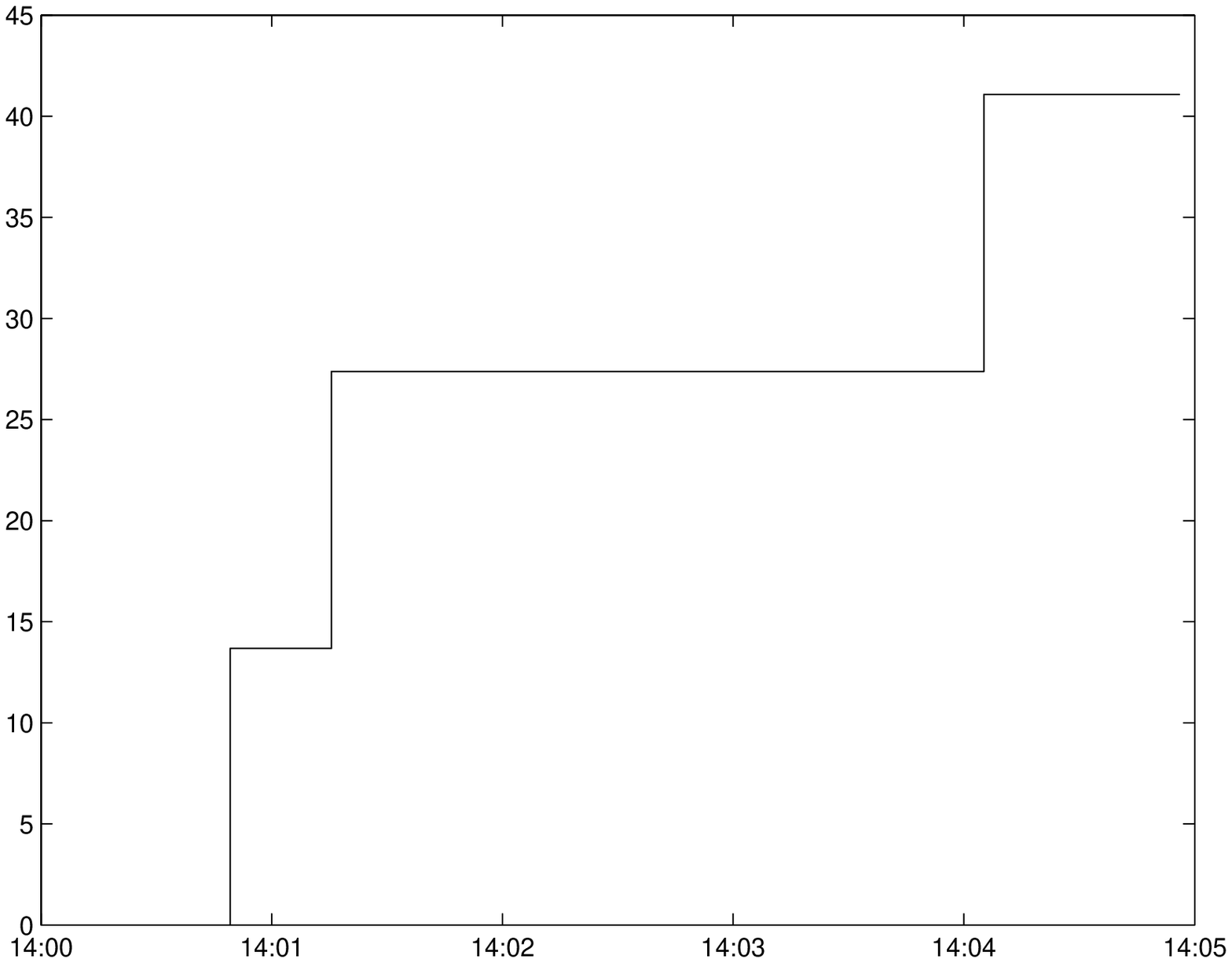}\\
\caption{Backtest example on AXA (November $5^{th}$ 2010). The strategy is used to sell a quantity of shares equal to 3 times the ATS within 5 minutes. Top: quotes of the trader (bold line), market best bid and ask quotes (thin lines). Trades are represented by dots. Bottom left: evolution of the inventory. Bottom right: cash at hand.}
\label{detail_5min_up}
\end{figure}

\begin{figure}[!h]
  \center
  \includegraphics[width=320pt]{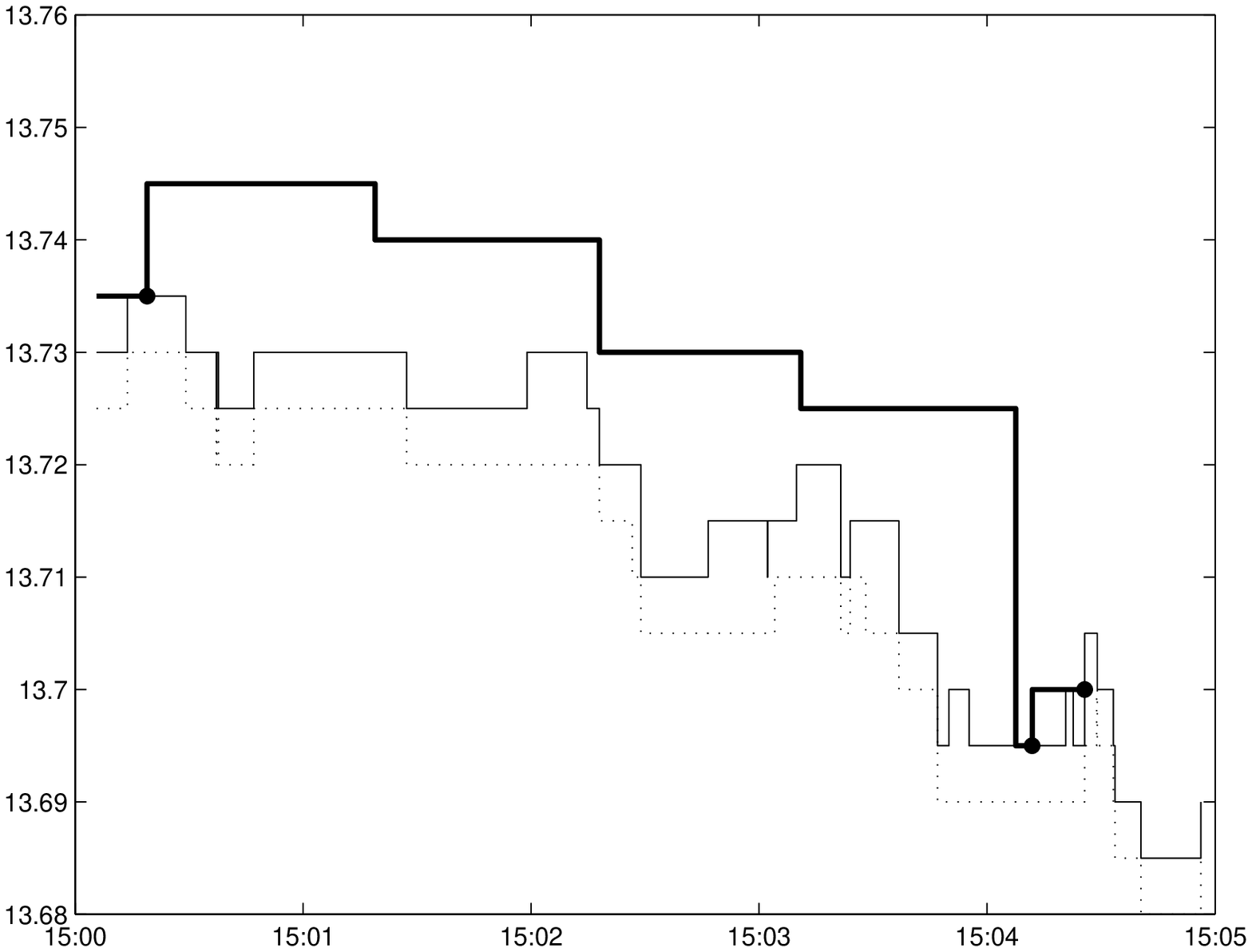}\\

  \includegraphics[width=180pt]{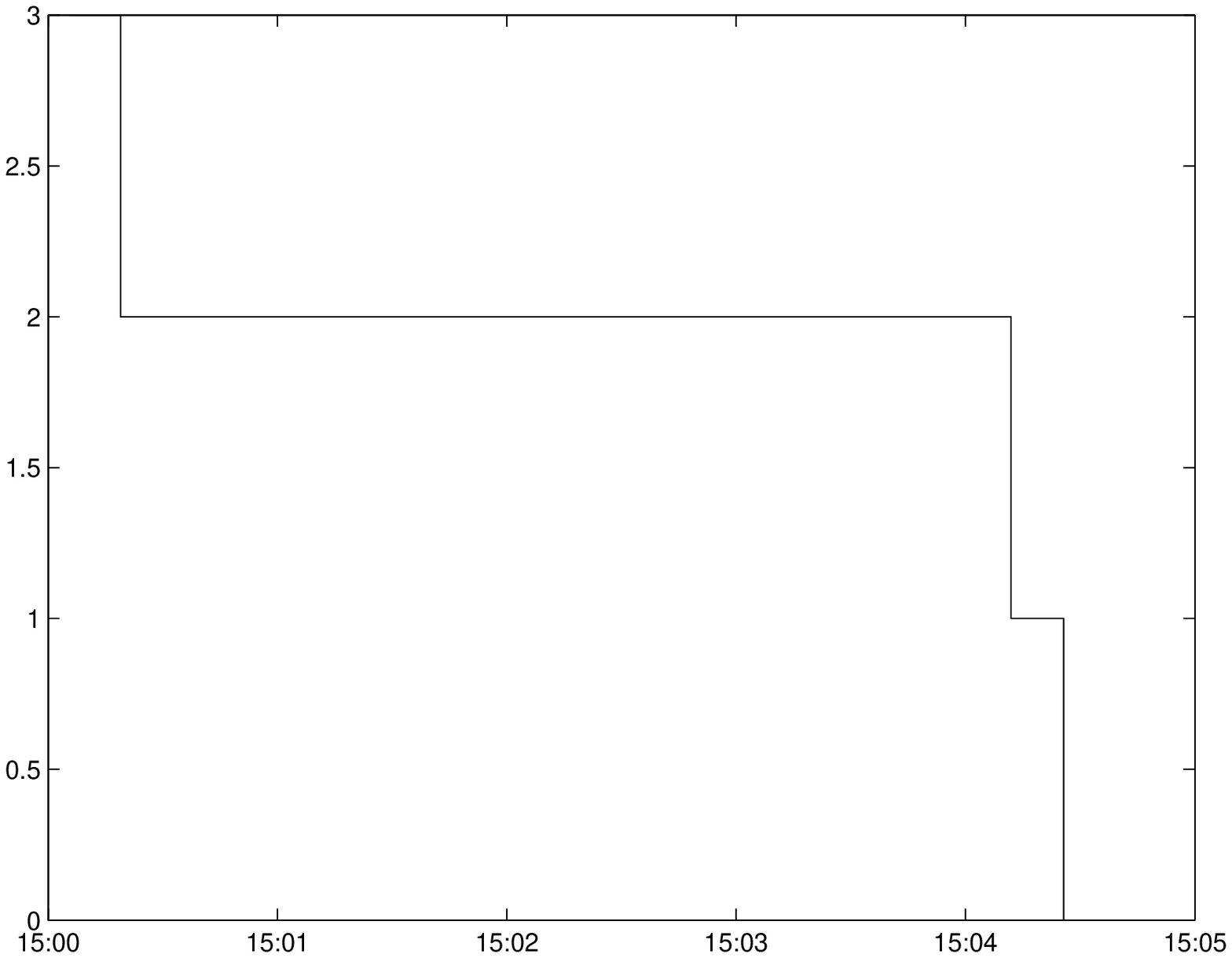}
  \hspace{1cm}
  \includegraphics[width=180pt]{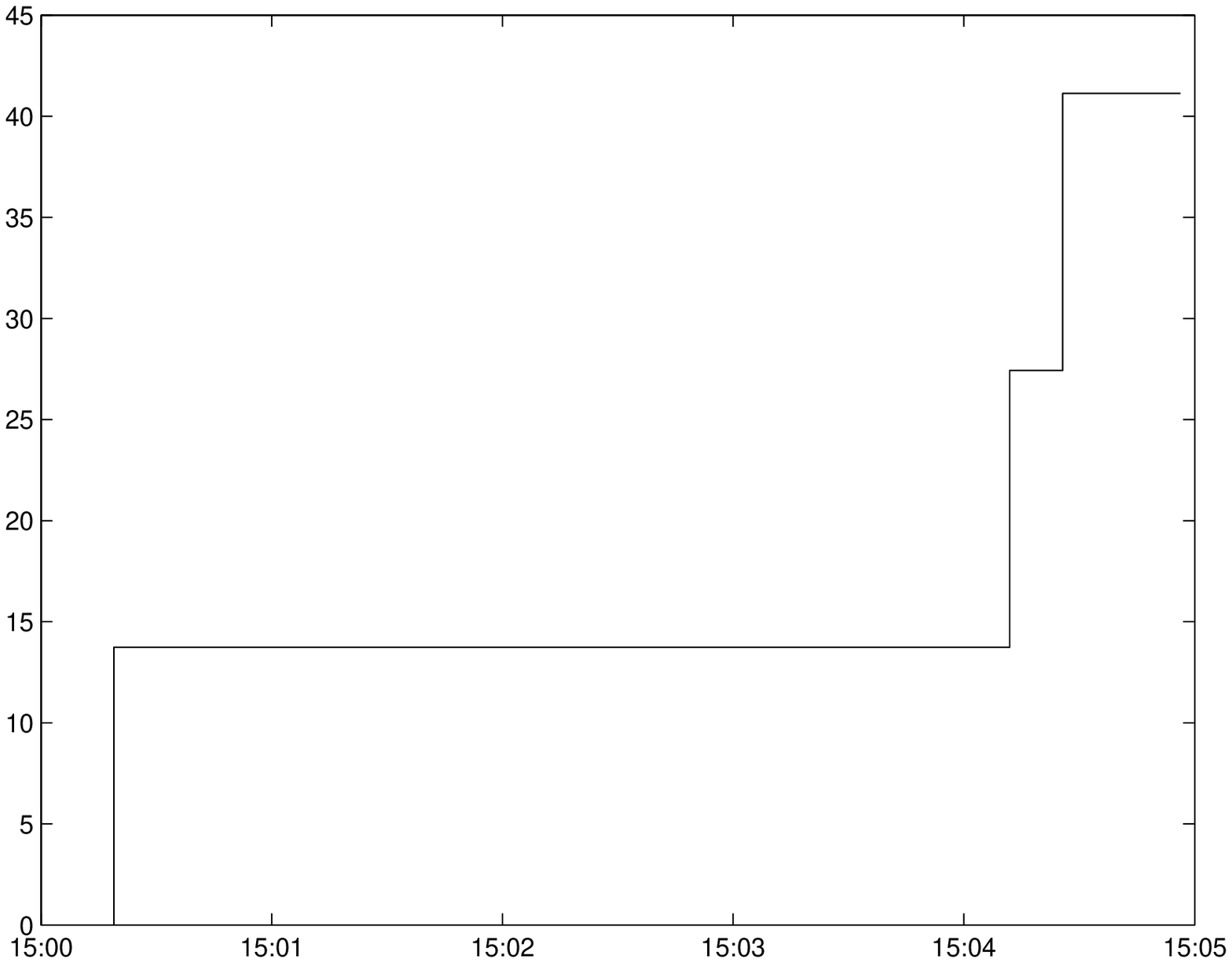}\\
\caption{Backtest example on AXA (November $5^{th}$ 2010). The strategy is used to sell a quantity of shares equal to 3 times the ATS within 5 minutes. Top: quotes of the trader (bold line), market best bid and ask quotes (thin lines). Trades are represented by dots. Bottom left: evolution of the inventory. Bottom right: cash at hand.}
\label{detail_5min_down}
\end{figure}

The first two examples (Figures~\ref{detail_5min_up} and \ref{detail_5min_down}) consist in liquidating a quantity of shares equal to 3 times the ATS\footnote{In the backtests we do not deal with quantity and priority issues in the order books and supposed that our orders were always entirely filled.}. The periods have been chosen to capture the behavior in both bullish and bearish markets.\\

On Figure~\ref{detail_5min_up}, we see that the first order is executed after 50 seconds. Then, since the trader has only 2 times the ATS left in his inventory, he sends an order at a higher price. Since the market price moves up, the second order is executed in the next 30 seconds, in advance on the average schedule. This is the reason why the trader places a new order far above the best ask. Since this order is not executed within the time window $\Delta t$, it is canceled and new orders are successively inserted with lower prices. The last trade happens less than 1 minute before the end of the period. Overall, on this example, the strategy works far better than a market order (even ignoring execution costs).\\

On Figure~\ref{detail_5min_down}, we see the use of the strategy in a bearish period. The first order is executed rapidly and since the market price goes down, the trader's last orders are only executed at the end of the period when prices of orders are lowered substantially as it becomes urgent to sell. Practically, this obviously raises the question of linking a trend detector to these optimal liquidation algorithms.\\
$$$$$$$$

\newpage

\begin{figure}[!h]
\center
  \includegraphics[width=350pt]{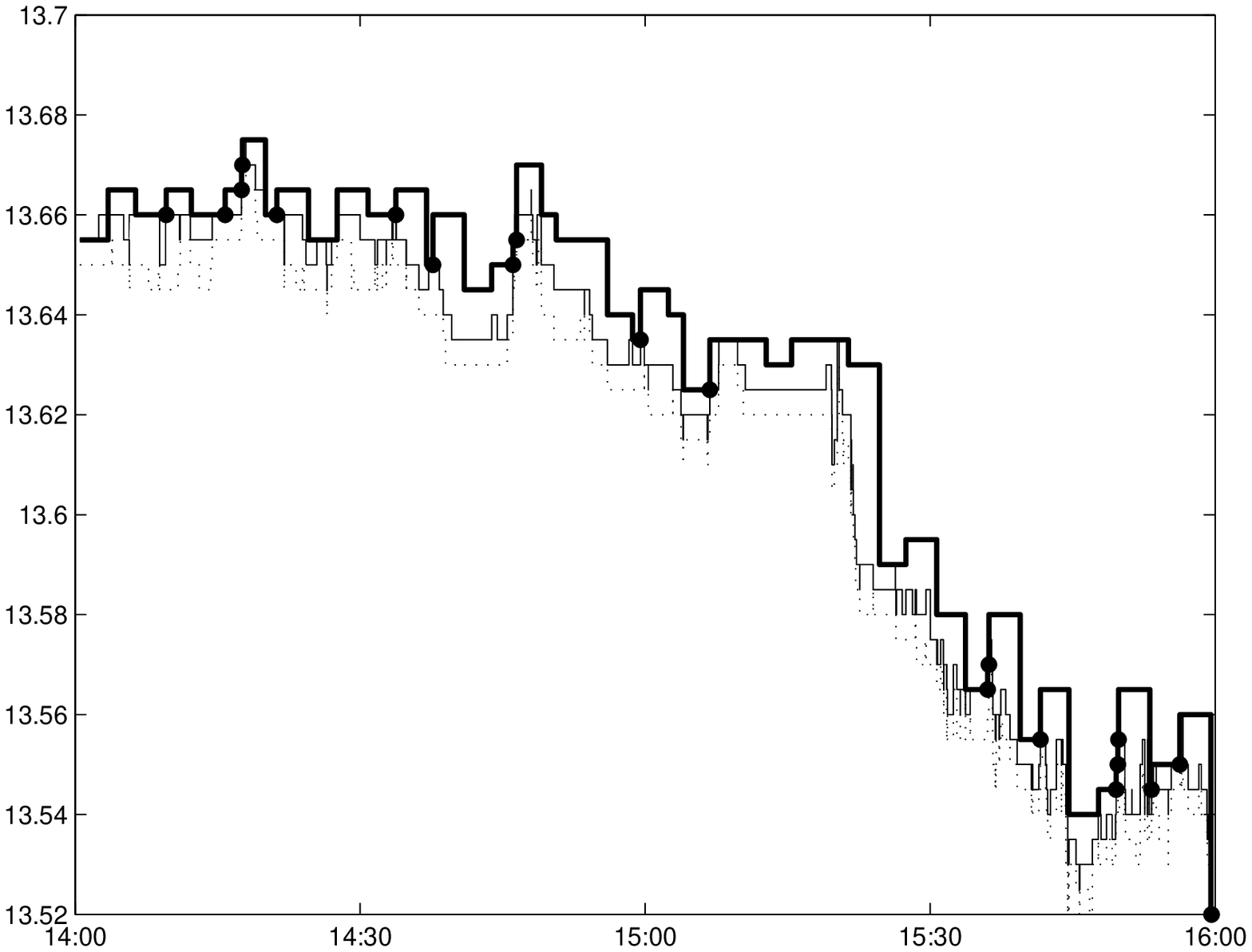}\\

  \includegraphics[width=180pt]{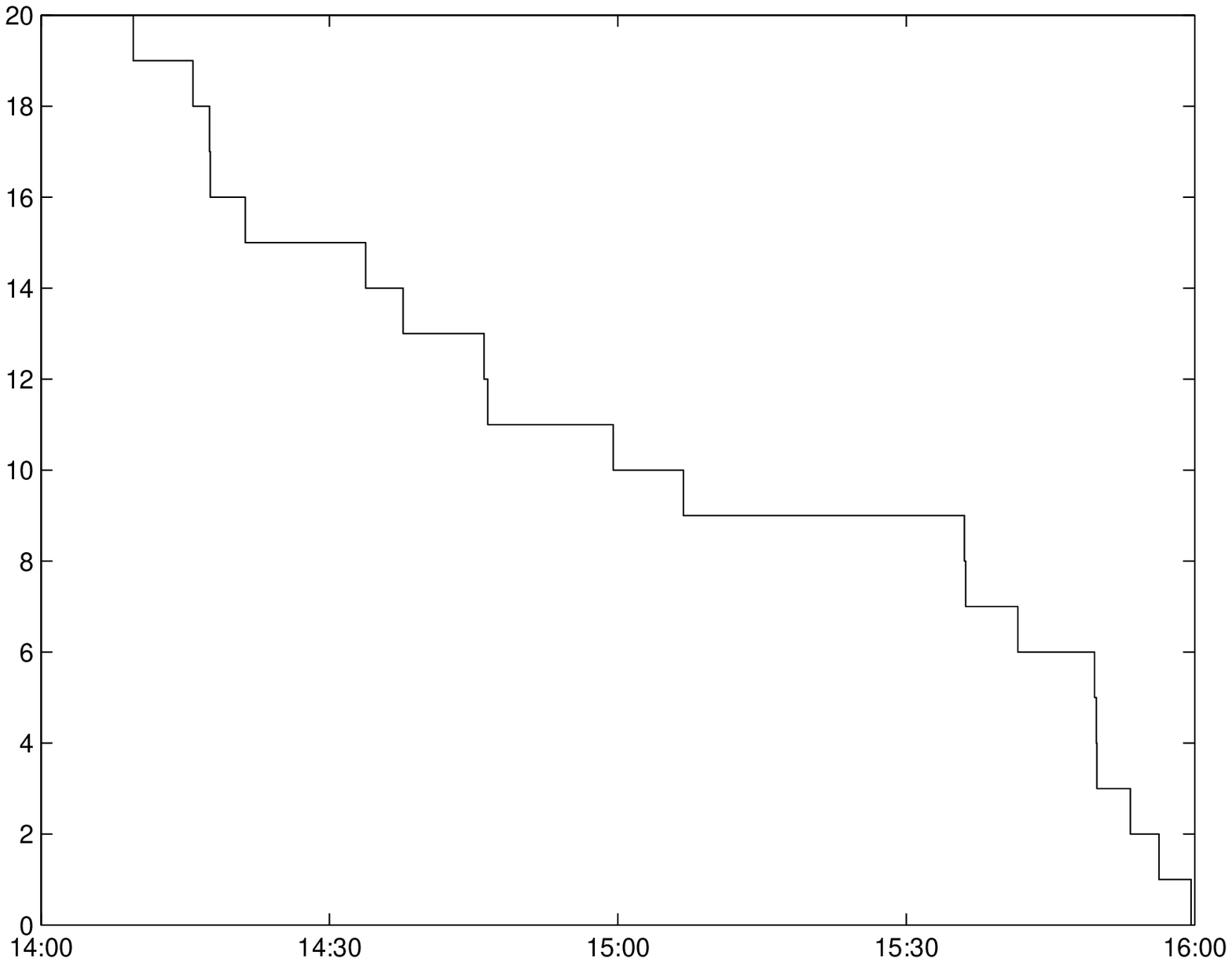}
  \hspace{1cm}
  \includegraphics[width=180pt]{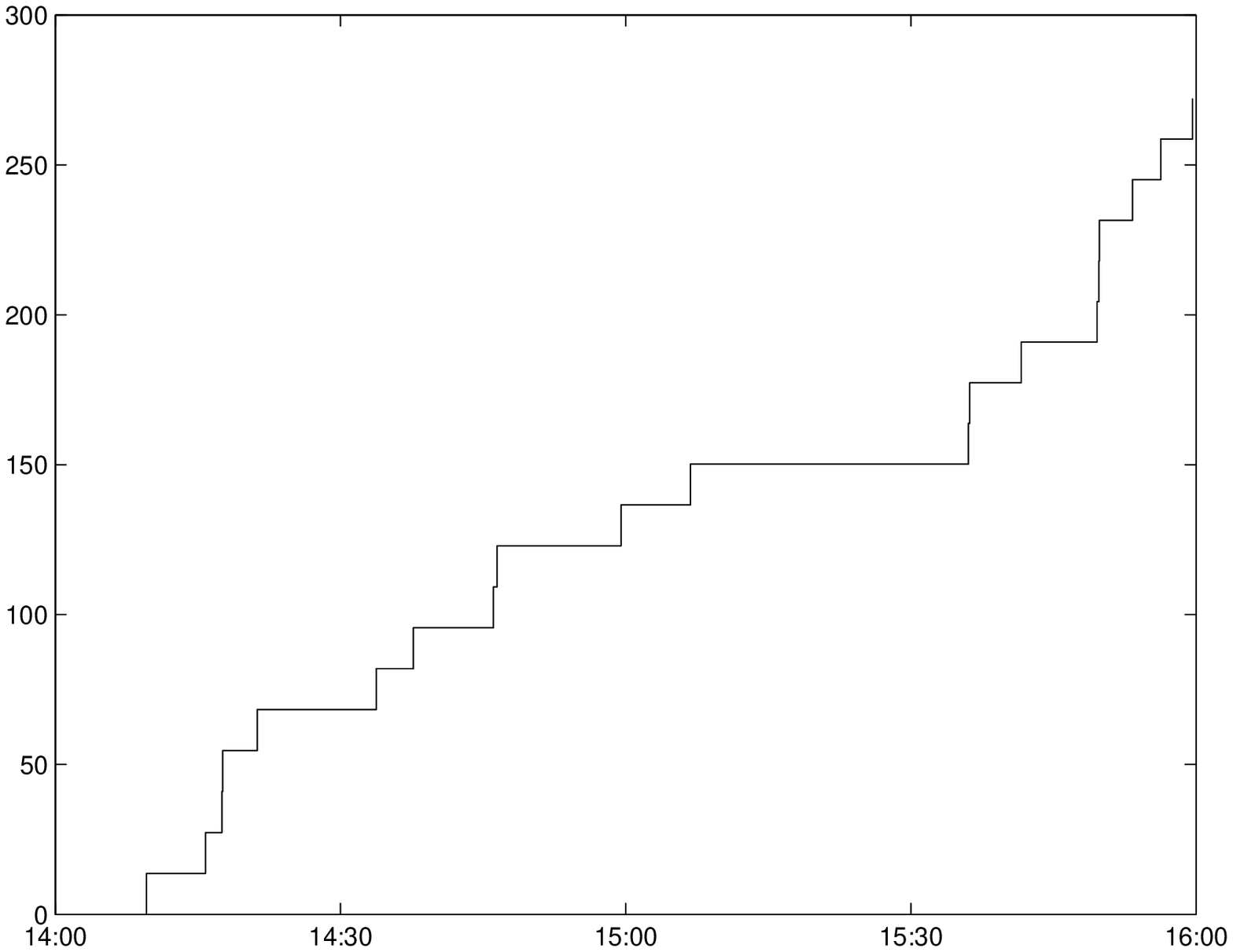}\\
\caption{Backtest example on AXA (November $8^{th}$ 2010). The strategy is used to sell a quantity of shares equal to 20 times the ATS within 2 hours. Top: quotes of the trader (bold line), market best bid and ask quotes (thin lines). Trades are represented by dots. Bottom left: evolution of the inventory. Bottom right: cash at hand.}
\label{detail_long}
\end{figure}

Finally, the model can also be used on longer periods and we exhibit the use of the algorithm on a period of two hours, to sell a quantity of shares equal to 20 times the ATS, representing here around 5\% of the volume during that period (Figure~\ref{detail_long}).
\vspace{2.5cm}

\newpage
\section*{Conclusion}

As claimed in the introduction, this paper is, to authors' knowledge, the first proposal to optimize the trade scheduling of large orders with small passive orders when price risk and non-execution risk are taken into account. The classical approach to optimal liquidation, following the Almgren-Chriss framework, consisted in a trade-off between price risk and execution cost/market impact. In the case of liquidity-providing orders, this trade-off disappears but a new risk is borne by the agent: non-execution risk.\\

The problem is then a new stochastic control problem and an innovative change of variables allows to reduce the 4-variable Hamilton-Jacobi-Bellman equation to a system of linear ordinary differential equations. Practically, the optimal quote can therefore be found in two steps: (1) solve a linear system of ODEs, (2) deduce the optimal price of the order to be sent to the market.\\

We studied various limiting cases that allowed to find the asymptotic behavior of the optimal strategy and to find the result obtained in parallel by Bayraktar and Ludkovski \cite{citeulike:9272222}, taking the risk-neutral limit. This also allowed us to confirm our intuition about the role played by the parameters.\\

Numerical experiments and backtests have been carried out and the results are promising. However, two possible improvements are worth the discussion.\\

First, no explicit model of what could be called ``passive market impact'' (\emph{i.e.} the perturbations of the price formation process by liquidity provision) is used here. Interestingly, Jaimungal, Cartea and Ricci \cite{cartea2011buy} recently introduced market impact in a similar model, the market impact occurring when execution takes place. We may consider introducing a similar effect in future versions of the model. Also, thanks to very promising and recent studies of the multi-dimensional point processes governing the arrival of orders (see for instance the link between the imbalance in the order flow and the moves of the price studied in \cite{citeulike:8531765} or \cite{citeulike:9249778}, or interesting properties of Hawkes-like models in \cite{citeulike:9217792}), we can hope for obtaining new models with passive market impact in the near future. The authors will try to embed them into the HJB framework used here.\\

Second, the separation of the variables $(x,s)$ and $(t,q)$ is a property associated to the use of a CARA utility function and to the brownian dynamic of the price and is independent of the exponential decay for the arrival of orders. An on-going work aims at generalizing the above model to a general function $\lambda^a(\cdot)$ using this separation of variables.\\

\newpage
\section*{Appendix}

\textbf{Proof of Proposition 1 and Theorem 1:}\\

First, let us remark that a solution $(w_q)_q$ of $(\mathcal{S})$ exists and is unique and that, by immediate induction, its components are strictly positive for all times. Then, let us introduce $u(t,x,q,s) = -\exp\left(-\gamma (x+qs)\right){w_q(t)}^{-\frac \gamma k}$.\\

We have:
$$\partial_t u + \mu \partial_s u + \frac 12 \sigma^2 \partial_{ss}^2 u =-\frac \gamma k \frac{\dot{w}_q(t)}{w_q(t)} u -\gamma q \mu u + \frac{\gamma^2\sigma^2}{2} q^2 u$$

Now, concerning the non-local part of the equation, we have:
$$\sup_{\delta^a} \lambda^a(\delta^a) \left[u(t,x+s + \delta^a,q-1,s) - u(t,x,q,s) \right]$$
$$=\sup_{\delta^a} A e^{-k\delta^a} u(t,x,q,s) \left[\exp\left(-\gamma \delta^a\right)\left(\frac{w_{q-1}(t)}{w_q(t)}\right)^{-\frac{\gamma}{k}} - 1 \right] $$

The first order condition of this problem corresponds to a maximum and writes:
$$(k+\gamma) \exp\left(-\gamma \delta^{a} \right)\left(\frac{w_{q-1}(t)}{w_q(t)}\right)^{-\frac{\gamma}{k}} = k$$

Hence we introduce the candidate $\delta^{a*}$ for the optimal control:

$$\delta^{a*} = \frac 1k \ln\left(\frac{w_{q}(t)}{w_{q-1}(t)}\right) + \frac 1\gamma \ln\left(1+\frac \gamma k\right)$$

and

$$\sup_{\delta^{a}} \lambda^a(\delta^{a}) \left[u(t,x+s+\delta^{a},q-1,s) - u(t,x,q,s) \right]$$$$ = -\frac{\gamma}{k+\gamma}A\exp(-k\delta^{a*})u(t,x,q,s)$$$$=-A\frac{\gamma}{k+\gamma} \left(1+\frac \gamma k\right)^{-\frac k \gamma} \frac{w_{q-1}(t)}{w_q(t)}  u(t,x,q,s)$$

Hence, putting the three terms together we get:
$$\partial_t u(t,x,q,s) + \mu \partial_s u(t,x,q,s) + \frac 12 \sigma^2 \partial_{ss}^2 u(t,x,q,s)$$$$ +\sup_{\delta^{a}} \lambda^a(\delta^{a}) \left[u(t,x+s+\delta^{a},q-1,s) - u(t,x,q,s) \right]$$
$$=-\frac \gamma k \frac{\dot{w}_q(t)}{w_q(t)} u - \gamma \mu q u + \frac{\gamma^2\sigma^2}{2} q^2 u -A\frac{\gamma}{k+\gamma} \left(1+\frac \gamma k\right)^{-\frac k \gamma} \frac{w_{q-1}(t)}{w_q(t)}  u $$
$$=-\frac \gamma k  \frac{u}{w_q(t)} \left[\dot{w}_q(t) + k\mu q w_q(t) - \frac{k\gamma\sigma^2}{2} q^2 w_q(t) + A\left(1+\frac \gamma k\right)^{-\left(1+\frac k \gamma\right)} w_{q-1}(t)  \right]=0$$

Now, noticing that the boundary and terminal conditions for $w_q$ are consistent with the conditions on $u$, we get that $u$ verifies $(\mathrm{HJB})$.\\

Now, we need to verify that $u$ is indeed the value function associated to the problem and to prove that our candidate $(\delta_t^{a*})_t$ is indeed the optimal control. To that purpose, let us consider a control $\nu \in \mathcal{A}$ and let us consider the following processes for $\tau \in [t,T]$:

$$dS^{t,s}_\tau = \mu d\tau + \sigma dW_\tau, \qquad S^{t,s}_t = s$$
$$dX^{t,x,\nu}_\tau = (S_\tau + \nu_\tau)  dN^a_\tau, \qquad  X^{t,x,\nu}_t = x$$
$$dq^{t,q,\nu}_\tau = - dN^a_\tau, \qquad  q^{t,q,\nu}_t = q$$
where the point process has stochastic intensity $(\lambda_\tau)_\tau$ with $\lambda_\tau = A e^{-k\nu_\tau} 1_{q_{\tau-} \ge 1}$\footnote{This intensity being bounded since $\nu$ is bounded from below.}.\\

Now, let us write Itô's formula for $u$ since $u$ is smooth:

$$u(T,X^{t,x,\nu}_{T-},q^{t,q,\nu}_{T-},S^{t,s}_{T}) = u(t,x,q,s)$$$$ + \int_t^T \left(\partial_\tau u(\tau,X^{t,x,\nu}_{\tau-},q^{t,q,\nu}_{\tau-},S^{t,s}_{\tau}) + \mu \partial_s u(\tau,X^{t,x,\nu}_{\tau-},q^{t,q,\nu}_{\tau-},S^{t,s}_\tau)  + \frac {\sigma^2}2 \partial^2_{ss} u(\tau,X^{t,x,\nu}_{\tau-},q^{t,q,\nu}_{\tau-},S^{t,s}_\tau)\right)d\tau$$
$$+ \int_t^T  \left(u(\tau,X^{t,x,\nu}_{\tau-}+S^{t,s}_\tau + \nu_\tau,q^{t,q,\nu}_{\tau-}-1,S^{t,s}_\tau) - u(\tau,X^{t,x,\nu}_{\tau-},q^{t,q,\nu}_{\tau-},S^{t,s}_\tau)\right) \lambda_\tau d\tau
$$$$  +\int_t^T \sigma \partial_s u(\tau,X^{t,x,\nu}_{\tau-},q^{t,q,\nu}_{\tau-},S^{t,s}_\tau) dW_{\tau}$$$$ + \int_t^T  \left(u(\tau,X^{t,x,\nu}_{\tau-}+S^{t,s}_\tau + \nu_\tau,q^{t,q,\nu}_{\tau-}-1,S^{t,s}_\tau) - u(\tau,X^{t,x,\nu}_{\tau-},q^{t,q,\nu}_{\tau-},S^{t,s}_\tau)\right) dM^a_\tau$$
where $M^a$ is the compensated process associated to $N^a$ for the intensity process $(\lambda_\tau)_\tau$.\\

Now, we have to ensure that the last two integrals consist of martingales so that their mean is $0$. To that purpose, let us notice that $\partial_s u = -\gamma q u$ and hence we just have to prove that:

$$\mathbb{E}\left[\int_t^T u(\tau,X^{t,x,\nu}_{\tau-},q^{t,q,\nu}_{\tau-},S^{t,s}_\tau)^2 d\tau\right] < +\infty$$

$$\mathbb{E}\left[\int_t^T \left|u(\tau,X^{t,x,\nu}_{\tau-}+S^{t,s}_\tau + \nu_\tau,q^{t,q,\nu}_{\tau-}-1,S^{t,s}_\tau)\right| \lambda_\tau d\tau\right] < +\infty$$
and
$$\mathbb{E}\left[\int_t^T \left|u(\tau,X^{t,x,\nu}_{\tau-},q^{t,q,\nu}_{\tau-},S^{t,s}_\tau)\right| \lambda_\tau d\tau\right] < +\infty$$
Now, remember that the process $q^{t,q,\nu}$ takes values between $0$ and $q$ and that $t \in [0,T]$. Hence, $\exists \varepsilon>0$, $w_q(t) > \varepsilon$ for the values of $t$ and $q$ under scrutiny and:

$$u(\tau,X^{t,x,\nu}_\tau,q^{t,q,\nu}_\tau,S^{t,s}_\tau)^2 \le \varepsilon^{-\frac{2\gamma}{k}}\exp{\left(-2\gamma(X^{t,x,\nu}_\tau+q^{t,q,\nu}_\tau S^{t,s}_\tau)\right)}$$
$$\le  \varepsilon^{-\frac{2\gamma}{k}}\exp{\left(-2\gamma(x-q\|\nu^-\|_{\infty} + 2q \inf_{\tau \in [t,T]} S^{t,s}_\tau 1_{\inf_{\tau \in [t,T]} S^{t,s}_\tau < 0} )\right)}$$
$$\le \varepsilon^{-\frac{2\gamma}{k}}\exp{\left({-2\gamma(x-q\|\nu^-\|_{\infty})}\right)} \left(1+\exp{\left(-2\gamma q\inf_{\tau \in [t,T]} S^{t,s}_\tau\right)} \right)$$
Hence:
$$\mathbb{E}\left[\int_t^T u(\tau,X^{t,x,\nu}_\tau,q^{t,q,\nu}_\tau,S^{t,s}_\tau)^2 d\tau\right] = \mathbb{E}\left[\int_t^T u(\tau,X^{t,x,\nu}_{\tau-},q^{t,q,\nu}_{\tau-},S^{t,s}_\tau)^2 d\tau\right]$$
$$\le \varepsilon^{-\frac{2\gamma}{k}}\exp{\left({-2\gamma(x-q\|\nu^-\|_{\infty})}\right)} (T-t) \left(1+\mathbb{E}\left[\exp{\left(-2\gamma q\inf_{\tau \in [t,T]} S^{t,s}_\tau\right)} \right]\right)$$
$$\le \varepsilon^{-\frac{2\gamma}{k}}\exp{\left({-2\gamma(x-q\|\nu^-\|_{\infty})}\right)} (T-t) \left(1+\mathbb{E}\left[\exp{\left(-2\gamma q\inf_{\tau \in [t,T]} S^{t,s}_\tau\right)} \right]\right)$$
$$\le \varepsilon^{-\frac{2\gamma}{k}}\exp{\left({-2\gamma(x-q\|\nu^-\|_{\infty})}\right)} (T-t) \left(1+e^{-2\gamma q s}\mathbb{E}\left[\exp{\left(2\gamma q \sigma\sqrt{T-t}|Y| \right)} \right]\right) < + \infty$$
where the last inequalities come from the reflection principle with $Y \sim \mathcal{N}(0,1)$ and the fact that $\mathbb{E}\left[e^{C|Y|}\right]<+\infty$ for any $C \in \mathbb{R}$.\\

Now, the same argument works for the second and third integrals, noticing that $\nu$ is bounded from below and that $\lambda$ is bounded.\\

Hence, since we have, by construction\footnote{This inequality is also true when the portfolio is empty because of the boundary conditions.}

$$\partial_\tau u(\tau,X^{t,x,\nu}_{\tau-},q^{t,q,\nu}_{\tau-},S^{t,s}_\tau) + \mu \partial_s u(\tau,X^{t,x,\nu}_{\tau-},q^{t,q,\nu}_{\tau-},S^{t,s}_\tau)  + \frac {\sigma^2}2 \partial^2_{ss} u(\tau,X^{t,x,\nu}_{\tau-},q^{t,q,\nu}_{\tau-},S^{t,s}_\tau)$$$$+ \left(u(\tau,X^{t,x,\nu}_{\tau-}+S^{t,s}_\tau + \nu_t,q^{t,q,\nu}_{\tau-}-1,S^{t,s}_\tau) - u(\tau,X^{t,x,\nu}_{\tau-},q^{t,q,\nu}_{\tau-},S^{t,s}_\tau)\right) \lambda_\tau \le 0$$

we obtain that

$$\mathbb{E}\left[u(T,X^{t,x,\nu}_T,q^{t,q,\nu}_T,S^{t,s}_T)\right] = \mathbb{E}\left[u(T,X^{t,x,\nu}_{T-},q^{t,q,\nu}_{T-},S^{t,s}_T)\right] \le u(t,x,q,s)$$
and this is true for all $\nu \in \mathcal{A}$. Since for $\nu = \delta^{a*}$ we have an equality in the above inequality we obtain that:

$$\sup_{\nu \in \mathcal{A}} \mathbb{E}\left[u(T,X^{t,x,\nu}_T,q^{t,q,\nu}_T,S^{t,s}_T)\right] \le u(t,x,q,s) = \mathbb{E}\left[u(T,X^{t,x,\delta^{a*}}_T,q^{t,q,\delta^{a*}}_T,S^{t,s}_T)\right]$$

This proves that $u$ is the value function and that $\delta^{a*}$ is optimal.\qed\\

\textbf{Proof of Proposition 2:}\\

We have that $$\forall q\in \mathbb{N}, \dot{w}_q(t) = (\alpha q^2 - \beta q) w_q(t) - \eta  w_{q-1}(t)$$

Hence if we consider for a given $Q \in \mathbb{N}$ the vector $ w(t) =
 \begin{pmatrix}
  w_{0}(t) \\
  w_{1}(t) \\
  \vdots  \\
  w_{Q}(t)
 \end{pmatrix}$ we have that $w'(t) = M w(t)$ where:

\[M=
 \begin{pmatrix}
  0 & 0                & \cdots & \cdots            & \cdots           & 0 \\
  -\eta & \alpha - \beta & 0      & \ddots            & \ddots           & \vdots \\
  0              &    \ddots           & \ddots          & \ddots           &   \ddots               & \vdots \\
    \vdots              &    \ddots           & \ddots          & \ddots           &   \ddots               & \vdots \\
                    \vdots& \ddots         & \ddots           & -\eta & \alpha (Q-1)^2 - \beta (Q-1) & 0   \\
 0                   & \cdots           & \cdots  & 0                & -\eta & \alpha Q^2 - \beta Q
\end{pmatrix}
\]
with $w(T) =  \begin{pmatrix}
  1 \\
  e^{-kb}\\
  \vdots  \\
  e^{-kbQ}
 \end{pmatrix}$. Hence we know that, if we consider a basis $(f_0, \ldots, f_{Q})$ of eigenvectors ($f_j$ being associated to the eigenvalue $\alpha j^2 - \beta j$), there exists $(c_0, \ldots, c_Q) \in \mathbb{R}^{Q+1}$ independent of $T$ such that:

$$w(t) = \sum_{j=0}^Q c_j e^{-(\alpha j^2 - \beta j)(T-t)} f_j$$

Consequently, since we assumed that $\alpha > \beta$, we have that $w^{\infty} := \lim_{T \to +\infty} w(0) = c_0 f_0$. Now, $w^{\infty}$ is characterized by:
$$ (\alpha q^2 - \beta q) w^{\infty}_q = \eta w^{\infty}_{q-1},  q>0 \qquad w^{\infty}_0 = 1 $$
As a consequence we have:
$$w^{\infty}_q = \frac{\eta^q}{q!} \prod_{j=1}^q \frac{1}{\alpha j - \beta}$$
The resulting asymptotic behavior for the optimal ask quote is:
$$\lim_{T\to +\infty} \delta^{a*}(0,q) = \frac{1}{k}\ln\left( \frac{A}{ 1+\frac{\gamma}{k}}\frac{1}{\alpha q^2 - \beta q} \right)$$
\qed\\

\textbf{Proof of Proposition 3:}\\

The result of Proposition 3 is obtained by induction. For $q=0$ the result is obvious.\\
Now, if the result is true for some $q$ we have that:

$$\dot{w}_{q+1}(t)= - \sum_{j=0}^q \frac{\eta^{j+1}}{j!} e^{-kb(q-j)}(T-t)^j$$
Hence:
$$w_{q+1}(t) = e^{-kb(q+1)} + \sum_{j=0}^q \frac{\eta^{j+1}}{(j+1)!} e^{-kb(q-j)}(T-t)^{j+1}$$
$$w_{q+1}(t) = e^{-kb(q+1)} + \sum_{j=1}^{q+1} \frac{\eta^{j}}{j!} e^{-kb(q-j+1)}(T-t)^{j}$$
$$w_{q+1}(t) = \sum_{j=0}^{q+1} \frac{\eta^{j}}{j!} e^{-kb(q+1-j)}(T-t)^{j}$$
This proves the results for $w$ and then the result follows for the optimal quote.\qed\\

\textbf{Proof of Proposition 4:}\\

Because the solutions depend continuously on $b$, we can directly get interested in the limiting equation:

$$\forall q\in \mathbb{N}, \dot{w}_q(t) = (\alpha q^2 - \beta q) w_q(t) - \eta  w_{q-1}(t)$$
with $w_q(T) = 1_{q=0}$ and $w_0 = 1$.\\

Then, if we define $v_q(t) = \lim_{b \to +\infty} \frac{w_{b,q}(t)}{A^q}$, $v$ solves:

$$\forall q\in \mathbb{N}, \dot{v}_q(t) = (\alpha q^2 - \beta q) v_q(t) - \tilde{\eta}  v_{q-1}(t)$$
with $v_q(T) = 1_{q=0}$ and $v_0 = 1$, where $\tilde{\eta}= \frac{\eta}{A}$ is independent of $A$.\\

Hence $v_q(t)$ is independent of $A$.\\

Now, for the trading intensity we have:

\begin{eqnarray*}
\lim_{b \to +\infty} A\exp{\left(-k\delta_b^{a*}(t,q) \right)} &=& \lim_{b \to +\infty} \frac{A w_{b,q-1}(t)}{w_{b,q}(t)}\left(1+\frac \gamma k\right)^{-\frac{k}{\gamma}}\\
 		    &=& \frac{v_{q-1}(t)}{v_{q}(t)}\left(1+\frac \gamma k\right)^{-\frac{k}{\gamma}}\\
\end{eqnarray*}
and this does not depend on $A$.\\

Eventually, since the limit of the trading intensity does not depend on $A$, the resulting trading curve does not depend on $A$ either.\qed\\

\textbf{Proof of Proposition 5:}\\

Using the preceding proposition, we can now reason in terms of $v$ and look for a solution of the form $v_q(t) = \frac{h(t)^q}{q!}$.\\
Then,
$$\forall q\in \mathbb{N}, \dot{v}_q(t) = - \beta q v_q(t) - \tilde{\eta}  v_{q-1}(t), \quad v_q(T) = 1_{q=0}, \quad v_0 = 1 $$
$$\iff h'(t) = -\beta h(t) - \tilde{\eta} \quad h(T) = 0$$

Hence, if $\beta = k\mu \neq 0$, the solution writes $v_q(t) = \frac{\tilde{\eta}^q}{q!} (\frac{\exp(\beta(T-t)) - 1}{\beta})^q$.

From Theorem 1, we obtain the limit of the optimal quote:

$$\lim_{b \to +\infty}\delta_b^{a*}(t,q) = \left( \frac{1}{k}\ln\left(\frac{\eta}{q} \frac{\exp(\beta(T-t)) - 1}{\beta} \right) + \frac 1\gamma \ln\left(1+\frac \gamma k\right)\right)$$

Using the expression for $\tilde{\eta}$, this can also be written:
$$\frac{1}{k}\ln\left( \frac{A}{ 1+\frac{\gamma}{k}  }\frac{1}{q}\frac{e^{\beta(T-t)} - 1}{\beta} \right)$$

Now, the for the trading intensity we get:

$$\lim_{b \to +\infty} A\exp{\left(-k\delta_b^{a*}(t,q) \right)} = \left(1+\frac \gamma k\right) q \frac{\beta}{e^{\beta(T-t)} - 1}$$

Hence, because the limit of the intensity is proportional to $q$, the limit $V(t)$ of the trading curve is characterized by the following ODE:

$$V'(t) = - \left(1+\frac \gamma k\right) V(t) \frac{\beta}{e^{\beta(T-t)} - 1}, \qquad V(0) = q_0$$

Solving this equation, we get:

\begin{eqnarray*}
V(t)& =& q_0 \exp\left(- \left(1+\frac \gamma k\right) \int_0^t { \frac{\beta}{e^{\beta(T-s)} - 1}} ds \right)\\
    & =& q_0 \exp\left(- \left(1+\frac \gamma k\right) \int_{e^{\beta(T-t)}}^{e^{\beta T}} \frac{1}{\xi(\xi-1)} d\xi\right)\\
    & =& q_0 \exp\left(- \left(1+\frac \gamma k\right) \left[\ln\left(1-\frac{1}{\xi}\right)\right]_{e^{\beta(T-t)}}^{e^{\beta T}} \right)\\
    & =& q_0 \left(\frac{1-e^{-\beta (T-t)}}{1-e^{-\beta T}}\right)^{1+\frac{\gamma}{k}}\\
\end{eqnarray*}

When $\beta= 0$ (\emph{i.e.} $\mu=0$) we proceed in the same way or by a continuity argument.\qed\\

\bibliographystyle{plain}

\begin{thebibliography}{10}

\bibitem{citeulike:6615020}
Aur\'{e}lien Alfonsi, Antje Fruth, and Alexander Schied.
\newblock {Optimal execution strategies in limit order books with general shape
  functions}.
\newblock {\em Quantitative Finance}, 10(2):143--157, 2010.

\bibitem{OPTEXECAC00}
R.~F. Almgren and N.~Chriss.
\newblock {Optimal execution of portfolio transactions}.
\newblock {\em Journal of Risk}, 3(2):5--39, 2000.

\bibitem{citeulike:5177342}
Robert Almgren.
\newblock {Optimal Trading in a Dynamic Market}.
\newblock Technical Report~2, 2009.

\bibitem{almgren03}
Robert~F. Almgren.
\newblock {Optimal execution with nonlinear impact functions and
  trading-enhanced risk}.
\newblock {\em Applied Mathematical Finance}, 10(1):1--18, 2003.

\bibitem{avst08}
Marco Avellaneda and Sasha Stoikov.
\newblock {High-frequency trading in a limit order book}.
\newblock {\em Quantitative Finance}, 8(3):217--224, 2008.

\bibitem{citeulike:9217792}
E.~Bacry, S.~Delattre, M.~Hoffmann, and J.~F. Muzy.
\newblock {Modeling microstructure noise with mutually exciting point
  processes}.
\newblock January 2011.

\bibitem{citeulike:9272222}
Erhan Bayraktar and Michael Ludkovski.
\newblock {Liquidation in Limit Order Books with Controlled Intensity}.
\newblock To appear in Mathematical Finance, 2012.

\bibitem{BLA98}
Dimitris Bertsimas and Andrew~W. Lo.
\newblock {Optimal control of execution costs}.
\newblock {\em Journal of Financial Markets}, 1(1):1--50, 1998.

\bibitem{citeulike:6607108}
Robert~J. Bloomfield, Maureen O'Hara, and Gideon Saar.
\newblock {The ``Make or Take'' Decision in an Electronic Market: Evidence on
  the Evolution of Liquidity}.
\newblock {\em Social Science Research Network Working Paper Series}, October
  2002.

\bibitem{citeulike:5797837}
Bruno Bouchard, Ngoc-Minh Dang, and Charles-Albert Lehalle.
\newblock {Optimal control of trading algorithms: a general impulse control
  approach}.
\newblock {\em SIAM J. Financial Mathematics}, 2011.

\bibitem{cartea2010modeling}
{\'A}.~Cartea and S.~Jaimungal.
\newblock Modeling asset prices for algorithmic and high frequency trading.
\newblock 2010.

\bibitem{cartea2012risk}
{\'A}.~Cartea and S.~Jaimungal.
\newblock Risk measures and fine tuning of high frequency trading strategies.
\newblock 2012.

\bibitem{cartea2011buy}
{\'A}.~Cartea, S.~Jaimungal, and J.~Ricci.
\newblock Buy low sell high: A high frequency trading perspective.
\newblock 2011.

\bibitem{citeulike:8531765}
Rama Cont and Adrien De~Larrard.
\newblock {Price Dynamics in a Markovian Limit Order Book Market}.
\newblock {\em Social Science Research Network Working Paper Series}, January
  2011.

\bibitem{citeulike:9249778}
Rama Cont, Arseniy Kukanov, and Sasha Stoikov.
\newblock {The Price Impact of Order Book Events}.
\newblock {\em Social Science Research Network Working Paper Series}, November
  2010.

\bibitem{NMD}
Ngoc-Minh Dang.
\newblock {Optimal trading with transient price impact. A comparison of
  discrete and continuous approach}.
\newblock {\em Preprint}, 2011.

\bibitem{citeulike:7369801}
P.~A. Forsyth, J.~S. Kennedy, S.~T. Tse, and H.~Windcliff.
\newblock {Optimal Trade Execution: A Mean-Quadratic-Variation Approach}, 2009.

\bibitem{citeulike:8043820}
Peter~A. Forsyth.
\newblock {A Hamilton Jacobi Bellman approach to optimal trade execution}.
\newblock {\em Applied Numerical Mathematics}, October 2010.

\bibitem{FOU06}
Thierry Foucault and Albert~J. Menkveld.
\newblock {Competition for Order Flow and Smart Order Routing Systems}.
\newblock October 2006.

\bibitem{citeulike:5177397}
Jim Gatheral.
\newblock {No-Dynamic-Arbitrage and Market Impact}.
\newblock {\em Social Science Research Network Working Paper Series}, October
  2008.

\bibitem{citeulike:6699563}
Jim Gatheral, Alexander Schied, and Alla Slynko.
\newblock {Transient Linear Price Impact and Fredholm Integral Equations}.
\newblock {\em Social Science Research Network Working Paper Series}, January
  2010.

\bibitem{citeulike:9272221}
Olivier Gu\'{e}ant, Charles-Albert Lehalle, and Joaquin Fernandez-Tapia.
\newblock {Dealing with the inventory risk}.
\newblock Technical report, 2011.

\bibitem{guilbaud2011optimal}
F.~Guilbaud and H.~Pham.
\newblock Optimal high frequency trading with limit and market orders.
\newblock 2011.

\bibitem{guilbaud2012optimal}
F.~Guilbaud and H.~Pham.
\newblock Optimal high frequency trading in a pro-rata microstructure with
  predictive information.
\newblock {\em Arxiv preprint arXiv:1205.3051}, 2012.

\bibitem{he2005dynamic}
H.~He and H.~Mamaysky.
\newblock Dynamic trading policies with price impact.
\newblock {\em Journal of Economic Dynamics and Control}, 29(5):891--930, 2005.

\bibitem{HoStoll}
Thomas Ho and Hans~R. Stoll.
\newblock {Optimal dealer pricing under transactions and return uncertainty}.
\newblock {\em Journal of Financial Economics}, 9(1):47--73, March 1981.

\bibitem{huberman2004price}
G.~Huberman and W.~Stanzl.
\newblock Price manipulation and quasi-arbitrage.
\newblock {\em Econometrica}, 72(4):1247--1275, 2004.

\bibitem{citeulike:2775239}
Gur Huberman and Werner Stanzl.
\newblock {Optimal Liquidity Trading}.
\newblock {\em Social Science Research Network Working Paper Series}, December
  2000.

\bibitem{kharroubi2009optimal}
I.~Kharroubi and H.~Pham.
\newblock Optimal portfolio liquidation with execution cost and risk.
\newblock {\em Arxiv preprint arXiv:0906.2565}, 2009.

\bibitem{citeulike:7358610}
Peter Kratz and Torsten Schöneborn.
\newblock {Optimal Liquidation in Dark Pools}.
\newblock {\em Social Science Research Network Working Paper Series}, February
  2009.

\bibitem{citeulike:5094012}
Charles-Albert Lehalle.
\newblock {Rigorous Strategic Trading: Balanced Portfolio and Mean-Reversion}.
\newblock {\em The Journal of Trading}, 4(3):40--46, 2009.

\bibitem{citeulike:5637814}
Charles-Albert Lehalle and Romain Burgot.
\newblock {The Established Liquidity Fragmentation Affects all Investors}.
\newblock Technical report, CA Cheuvreux, March 2009.

\bibitem{citeulike:7621540}
Charles-Albert Lehalle, Olivier Gu\'{e}ant, and Julien Razafinimanana.
\newblock {High Frequency Simulations of an Order Book: a Two-Scales Approach}.
\newblock In F.~Abergel, B.~K. Chakrabarti, A.~Chakraborti, and M.~Mitra,
  editors, {\em Econophysics of Order-Driven Markets}, New Economic Windows.
  Springer, 2010.

\bibitem{lorenz2011mean}
J.~Lorenz and R.~Almgren.
\newblock Mean--variance optimal adaptive execution.
\newblock {\em Applied Mathematical Finance}, 2011.

\bibitem{ANYA05}
Anna Obizhaeva and Jiang Wang.
\newblock {Optimal Trading Strategy and Supply/Demand Dynamics}.
\newblock {\em Social Science Research Network Working Paper Series}, February
  2005.

\bibitem{citeulike:8531791}
S.~Predoiu, G.~Shaikhet, and S.~Shreve.
\newblock {Optimal Execution of a General One-Sided Limit-Order Book}.
\newblock Technical report, Carnegie-Mellon University, September 2010.

\bibitem{schied2009risk}
A.~Schied and T.~Schöneborn.
\newblock Risk aversion and the dynamics of optimal liquidation strategies in
  illiquid markets.
\newblock {\em Finance and Stochastics}, 13(2):181--204, 2009.

\end{thebibliography}

\end{document}